\documentclass[journal]{IEEEtran}

\usepackage{amsfonts}
\usepackage{mathrsfs}
\usepackage{epsf,graphics,latexsym,graphicx}
\usepackage{amsmath} % assumes amsmath package installed
\usepackage{amssymb}  % assumes amsmath package installed
\usepackage{url}
\usepackage{color}

\makeatletter
\def\ps@headings{%
\def\@oddhead{\mbox{}\scriptsize\rightmark \hfil \thepage}%
\def\@evenhead{\scriptsize\thepage \hfil \leftmark\mbox{}}%
\def\@oddfoot{}%
\def\@evenfoot{}}
\makeatother

\newtheorem{theorem}{Theorem}

\newcommand{\cG}{\mathcal{G}}
\newcommand{\cB}{\mathcal{B}}
\newcommand{\cD}{\mathcal{D}}
\newcommand{\cL}{\mathcal{L}}
\newcommand{\cN}{\mathcal{N}}
\newcommand{\cF}{\mathcal{F}}

\newcommand{\bE}{\mathbb{E}}

\begin{document}

\title{Backpressure-based Packet-by-Packet Adaptive Routing in Communication Networks}

\author{Eleftheria Athanasopoulou, Loc Bui, Tianxiong Ji, R. Srikant, and Alexander Stoylar}
%\author{Eleftheria Athanasopoulou{$^\dagger$}, Loc Bui{$^\dagger$}, Tianxiong Ji{$^\dagger$}, R. Srikant{$^\dagger$}, and Alexander Stoylar{$^\ddagger$} \\
%Coordinated Science Laboratory and\\
%Department of Electrical and Computer Engineering\\
%University of Illinois at Urbana-Champaign{$^\dagger$}\\
%Bell Labs, Alcatel-Lucent, NJ, USA{$^\ddagger$}\\
%Email: \{athanaso, locbui, tji2,
%rsrikant\}@illinois.edu{$^\dagger$},
%stolyar@research.bell-labs.com{$^\ddagger$}
%}

\maketitle

\begin{abstract}
Backpressure-based adaptive routing algorithms where each packet is
routed along a possibly different path have been extensively studied
in the literature. However, such algorithms typically result in poor
delay performance and involve high implementation complexity. In
this paper, we develop a new adaptive routing algorithm built upon
the widely-studied back-pressure algorithm. We decouple the routing
and scheduling components of the algorithm by designing a
probabilistic routing table which is used to route packets to
per-destination queues. The scheduling decisions in the case of
wireless networks are made using counters called shadow queues. The
results are also extended to the case of networks which employ
simple forms of network coding. In that case, our algorithm provides
a low-complexity solution to optimally exploit the routing-coding
tradeoff.
\end{abstract}

\section{Introduction}
\label{sec:intro}

\footnotetext[1]{This work was supported by MURI BAA 07-036.18, ARO
MURI, DTRA Grant HDTRA1-08-1-0016, and NSF grants 07-21286, 05-19691
and 03-25673.
\\ \indent E. Athanasopoulou, T. Ji and R. Srikant are with Coordinated Science Laboratory and
Department of Electrical and Computer Engineering, University of Illinois at Urbana-Champaign. (email: \{athanaso, tji2, rsrikant\}@illinois.edu).
\\ \indent L. Bui is with Airvana Inc., USA. (email: locbui@ieee.org).
\\ \indent A. Stolyar is with Bell Labs, Alcatel-Lucent, NJ, USA. (email: stolyar@research.bell-labs.com).

}

The back-pressure algorithm introduced in \cite{taseph92} has been
widely studied in the literature. While the ideas behind scheduling
using the weights suggested in that paper have been successful in
practice in base stations and routers, the adaptive routing
algorithm is rarely used. The main reason for this is that the
routing algorithm can lead to poor delay performance due to routing
loops. Additionally, the implementation of the back-pressure
algorithm requires each node to maintain per-destination queues
which can be burdensome for a wireline or wireless router. Motivated
by these considerations, we reexamine the back-pressure routing
algorithm in the paper and design a new algorithm which has much
superior performance and low implementation complexity.

Prior work in this area \cite{neemodroh05} has recognized the
importance of doing shortest-path routing to improve delay
performance and modified the back-pressure algorithm to bias it
towards taking shortest-hop routes.  A part of our algorithm has
similar motivating ideas, but we do much more. In addition to
provably throughput-optimal routing which minimizes the number of
hops taken by packets in the network, we decouple routing and
scheduling in the network through the use of probabilistic routing
tables and the so-called shadow queues. The min-hop routing idea was
studied first in a conference paper \cite{buisristo09} and shadow
queues were introduced in \cite{buisristo08}, but the key step of
decoupling the routing and scheduling which leads to both dramatic
delay reduction and the use of per-next-hop queueing is original
here. The min-hop routing idea is also studied in \cite{yinshared09}
but their solution requires even more queues than the original
back-pressure algorithm.

We also consider networks where simple forms of network coding is
allowed \cite{katrahhukatmedcro06}. In such networks, a relay
between two other nodes XORs
packets and broadcast them to decrease the number of transmissions.
There is a tradeoff between choosing long routes
to possibly  increase network coding opportunities (see the notion
of reverse carpooling in \cite{effhokim06}) and choosing short
routes to reduce resource usage. Our adaptive routing
algorithm can be modified to automatically realize this tradeoff
with good delay performance. In addition, network coding requires
each node to maintain more queues \cite{sefmarkoz09} and our routing
solution at least reduces the number of queues to be maintained for
routing purposes, thus partially mitigating the problem. An offline
algorithm for optimally computing the routing-coding tradeoff was
proposed in \cite{senrayban07}. Our optimization formulation
bears similarities to this work but our main focus is on designing
low-delay on-line algorithms. Back-pressure solutions to network
coding problems have also been studied in \cite{hovis09, erylun07,
cheholowchidoy07}, but the adaptive routing-coding tradeoff solution
that we propose here has not been studied previously.

We summarize our main results below.
\begin{itemize}

\item
Using the concept of shadow queues, we decouple routing and
scheduling. A shadow network is used to update a probabilistic
routing table which packets use upon arrival at a node. The
back-pressure-based scheduling algorithm is used to serve FIFO
queues over each link.

\item
The routing algorithm is designed to minimize the average number of
hops used by packets in the network. This idea, along with the scheduling/routing decoupling,
leads to delay reduction compared with the traditional back-pressure algorithm.

\item
Each node has to maintain counters, called shadow queues, per
destination. This is very similar to the idea of maintaining a
routing table per destination. But the real queues at each node are
per-next-hop queues in the case of networks which do not employ
network coding. When network coding is employed, per-previous-hop
queues may also be necessary but this is a requirement imposed by
network coding, not by our algorithm.

\item
The algorithm can be applied to wireline and wireless networks.
Extensive simulations show dramatic improvement in delay performance
compared to the back-pressure algorithm.

\end{itemize}

The rest of the paper is organized as follows. We present the network model in Section \ref{sec:network-model}. In Section \ref{sec:back-pressure} and \ref{sec:m-algorithm}, the traditional back-pressure algorithm and its modified version are introduced. We develop our adaptive routing and scheduling algorithm for wireline and wireless networks with and without network coding in Section \ref{sec:adaptive-routing}, \ref{sec:implement_detail} and \ref{sec:extend-network-coding}. In Section \ref{sec:simulation}, the simulation results are presented. We conclude our paper in Section \ref{sec:conclusion}.

\section{The Network Model}
\label{sec:network-model}
We consider a multi-hop wireline or wireless network represented by a directed graph $\cG=(\cN,\cL),$ where $\cN$ is the set of nodes and $\cL$ is the set of directed links. A directed link that can transmit packets from node $n$ to node $j$ is denoted by $(nj)\in \cL.$ We assume that time is slotted and define the link capacity $c_{nj}$ to be the maximum number of packets that link $(nj)$ can transmit in one time slot.

Let $\cF$ be the set of flows that share the network. Each flow is associated with a source node and a destination node, but no route is specified between these nodes. This means that the route can be quite different for packets of the same flow. Let $b(f)$ and $e(f)$ be source and destination nodes, respectively, of flow $f.$ Let $x_f$ be the rate (packets/slot) at which packets are generated by flow $f.$ If the demand on the network, i.e., the set of flow rates, can be satisfied by the available capacity, there must exist a routing algorithm and a scheduling algorithm such that the link rates lie in the capacity region. To precisely state this condition, we define $\mu_{nj}^{d}$ to be the rate allocated on link $(nj)$ to packets destined for node $d.$  Thus, the total rate allocated to all flows at link $(nj)$ is given by $\mu_{nj}:= \displaystyle \sum_{d \in \cN} \mu_{nj}^d.$ Clearly, for the network to be able to meet the traffic demand, we should have:\\
$$\{\mu_{nj}\}_{(nj) \in \cL} \in \Lambda,$$ where $\Lambda$ is the capacity region of the network for $1$-hop traffic. The capacity region of the network for $1$-hop traffic contains all sets of rates that are stabilizable by some kind of scheduling policy assuming all traffics are 1-hop traffic.
As a special case, in the wireline network, the constraints are:
$$\mu_{nj} \leq c_{nj},\qquad \forall (nj).$$
As opposed to $\Lambda,$ let $\Upsilon$ denote the capacity region
of the multi-hop network, i.e., for any set of flows $\{x_f\}_{f \in
\cF} \in \Upsilon,$ there exists some routing and scheduling
algorithms that stabilize the network.

In addition, a flow conservation constraint must be satisfied at
each node, i.e., the total rate at which traffic can possibly arrive
at each node destined to $d$ must be less than or equal to the total
rate at which traffic can depart from the node destined to $d:$
\begin{eqnarray} \label{eq: Wireline Constraints}
\begin{array}{l}
\displaystyle \displaystyle \sum_{f \in \cF} x_f I_{\{
b(f)=n, e(f)=d \}}
+ \displaystyle \sum_{l:(ln) \in \cL} \mu_{ln}^d \\
\;\;\;\; \leq \displaystyle  \sum_{j: (nj) \in \cL}
\mu_{nj}^d,
\end{array} \end{eqnarray}
where $I$ denotes the indicator function.
Given a set of arrival rates $x=\{x_f\}_{f\in \cF}$ that can be
accommodated by the network, one version of the multi-commodity flow
problem is to find the traffic splits $\mu_{nj}^d$ such that
(\ref{eq: Wireline Constraints}) is satisfied.  However, finding the
appropriate traffic split is computationally prohibitive and
requires knowledge of the arrival rates. The back-pressure algorithm
to be described next is an adaptive solution to the multi-commodity
flow problem.

\section{Throughput-Optimal Back-pressure Algorithm and Its Limitations}
\label{sec:back-pressure} The back-pressure algorithm was first
described in \cite{taseph92} in the context of wireless networks and
independently discovered later in \cite{awelei93} as a
low-complexity solution to certain multi-commodity flow problems.
This algorithm combines the scheduling and routing functions
together. While many variations of this basic algorithm have been
studied, they primarily focus on maximizing throughput and do not
consider QoS performance. Our algorithm uses some of these ideas as
building blocks and therefore, we first describe the basic
algorithm, its drawbacks and some prior solutions.

The algorithm maintains a queue for each destination at each node.
Since the number of destinations can be as large as the number of
nodes, this per-destination queueing requirement can be quite large
for practical implementation in a network. At each link, the
algorithm assigns a weight to each possible destination which is
called {\it back-pressure.} Define the back-pressure at link $(nj)$
for destination $d$ at slot $t$ to be
\begin{eqnarray*}\label{eq:back-pressure}
w_{nj}^d[t] = Q_{nd}[t] - Q_{jd}[t],
\end{eqnarray*}
where $Q_{nd}[t]$ denotes the number of packets at node $n$ destined
for node $d$ at the beginning of time slot $t.$ Under this notation,
$Q_{nn}[t]=0, \forall t.$ Assign a weight $w_{nj}$ to each link
$(nj),$ where $w_{nj}$ is defined to be the maximum back-pressure
over all possible destinations, i.e.,
\begin{eqnarray*}\label{eq:weight}
w_{nj}[t]=\max_{d} w_{nj}^d[t].
\end{eqnarray*}
Let $d^*_{nj}$ be the destination which has the maximum weight on
link $(nj),$
\begin{eqnarray}
d^*_{nj}[t]  = \displaystyle  \arg \max_d \{ w_{nj}^d[t] \}.
\end{eqnarray}
If there are ties in the weights, they can be broken arbitrarily.
Packets belonging to destination $d_{nj}^*[t]$ are scheduled for
transmission over the activated link $(nj).$ A schedule is a set of
links that can be activated simultaneously without interfering with each
other. Let $\Gamma$ denote the set of all schedules. The
back-pressure algorithm finds an optimal schedule $\pi^*[t]$ which
is derived from the optimization problem:
\begin{eqnarray}
\pi^*[t]=\displaystyle \arg \max_{\pi \in \Gamma}  \sum_{(nj) \in \pi}c_{nj} w_{nj}[t] .
\end{eqnarray}
Specially, if the capacity of every link has the same value, the
chosen schedule maximizes the sum of weights in any schedule.

At time $t,$ for each activated link $(nj) \in \pi^*[t]$ we remove
$c_{nj}$ packets from $Q_{nd_{nj}^*[t]}$ if possible, and transmit
those packets to $Q_{jd_{nj}^*[t]}.$ We assume that the departures
occur first in a time slot, and external arrivals and packets
transmitted over a link $(nj)$ in a particular time slot are
available to node $j$ at the next time slot. Thus the evolution of
the queue $Q_{nd}[t]$ is as follows:
\begin{eqnarray}\label{eq: queue dynamics}
\begin{array}{lll} Q_{nd}[t+1] &=&
Q_{nd}[t]
- \displaystyle \sum_{j: (nj) \in \cL}
I_{\{d^*_{nj}[t]=d\}}\,\hat{\mu}_{nj}[t]
\\&&+ \displaystyle \sum_{l: (ln) \in \cL} I_{\{d^*_{ln}[t]=d\}}
\,\hat{\mu}_{ln}[t]
\\&&+\displaystyle\sum_{f \in \cF}
I_{\{b(f)=n,e(f)=d\}}\,a_f[t],
\end{array}
\end{eqnarray}
where $\hat{\mu}_{nj}[t]$ is the number of packets transmitted over
link $(nj)$ in time slot $t$ and $a_f[t]$ is the number of packets
generated by flow $f$ at time $t.$
It has been shown in \cite{taseph92} that the back-pressure
algorithm maximizes the throughput of the network.

A key feature of the back-pressure algorithm is that packets may not
be transferred over a link unless the back-pressure over a link is
non-negative and the link is included in the picked schedule. This
feature prevents further congesting nodes that are already
congested, thus providing the adaptivity of the algorithm. Notice
that because all links can be activated without interfering with
each other in the wireline network, $\Gamma$ is the set of all
links. Thus the back-pressure algorithm can be localized at each
node and operated in a distributed manner in the wireline network.

The back-pressure algorithm has several disadvantages that prohibit
practical implementation:
\begin{itemize}

\item The back-pressure algorithm requires maintaining queues for each potential destination at each node.
This queue management requirement could be a prohibitive overhead
for a large network.

\item The back-pressure algorithm is an adaptive routing algorithm which explores the network resources and adapts to different levels of traffic intensity.
However it might also lead to high delays because it may choose long
paths unnecessarily. High delays are also a result of maintaining a
large number of queues at each node. Only one queue can be scheduled
at a time, and the unused service could further contribute to high
latency.

\end{itemize}

In this paper, we address the high delay and queueing complexity
issues. The computational complexity issue for wireless networks is not addressed here. We
simply use the recently studied greedy maximal scheduling (GMS)
algorithm. Here we call it the {\it largest-weight-first} algorithm,
in short, LWF algorithm. LWF algorithm requires the same queue
structure that the back-pressure algorithm uses. It also calculates
the back-pressure at each link using the same way. The difference
between these two algorithms only lies in the methods to pick a
schedule. Let $\mathcal S$ denote the set of all links initially.
Let $\mathcal N_b (l)$ be the set of links within the interference
range of link $l$ including $l$ itself. At each time slot, the LWF
algorithm picks a link $l$ with the maximum weight first, and
removes links within the interference range of link $l$ from
$\mathcal S,$ i.e., $\mathcal S = \mathcal S \backslash \mathcal
N_b(l)$; then it picks the link with the maximum weight in the
updated set $\mathcal S,$ and so forth. It should be noticed that
LWF algorithm reduces the computational complexity with a price of
the reduction of the network capacity region. The LWF algorithm
where the weights are queue lengths (not back-pressures) has been
extensively studied in \cite{dimwal06, joolinshr08, BZM06,
lecnisri09, liboyxia09}. While these studies indicate that there may
be reduction in throughput due to LWF in certain special network
topologies, it seems to perform well in simulations and so we adopt
it here.

In the rest of the paper, we present our main results which
eliminate many of the problems associated with the back-pressure
algorithm.

\section{Min-resource Routing Using Back-pressure Algorithm}
\label{sec:m-algorithm}
As mentioned in Section \ref{sec:back-pressure}, the back-pressure algorithm explores all paths in the network and as a result may choose paths which are unnecessarily long which may even contain loops, thus leading to poor performance. We address this problem by introducing a cost function which measures the total amount of resources used by all flows in the network. Specially, we add up traffic loads on all links in the network and use this as our cost function. The goal then is to minimize this cost subject to network capacity constraints.

Given a set of packet arrival rates that lie within the
capacity region, our goal is to find the routes for flows so
that we use as few resources as possible in the network. Thus, we
formulate the following optimization problem:
\begin{eqnarray}\label{eqn:minhop_opt}
&\min& \sum_{(nj)\in \cL} \mu_{nj} \\
&s.t.& \sum_{f\in \cF}x_f I_{\{b(f)=n,e(f)=d\}} + \sum_{(ln)\in \cL} \mu^d_{ln}
~\leq~ \sum_{(nj)\in \cL} \mu^d_{nj} , \nonumber\\
&& \qquad \qquad \qquad \qquad \qquad \quad
\forall d\in \cN , n \in \cN, \nonumber\\
&& \{\mu_{nj}\}_{(nj)\in  \cL} ~\in~ \Lambda. \nonumber
\end{eqnarray}

%An algorithm that asymptotically solves the min-resource
%routing problem (\ref{eqn:minhop_opt}) is as follows. (It is a
%special case of the algorithm in \cite{sto05}, where the
%scaling parameter $1/M$ is called $\beta$.)

We now show how a modification of the back-pressure algorithm can be used to solve this min-resource routing problem. (Note that similar approaches have been used in \cite{linshr04, neemodli05, sto05, erysri05, erysri06} to solve related resource allocation problems.)

%We now show how a modification of the back-pressure algorithm can be used to solve this min-resource routing problem \cite{linshr04, neemodli05, sto05, erysri05, erysri06}.

Let $\{q_{nd}\}$ be the Lagrange multipliers corresponding to the flow conservation constraints in problem (\ref{eqn:minhop_opt}). Appending these constraints to the objective, we get
\begin{eqnarray}\label{eqn:solve_min_opt}
& &\min_{\mbox{{\boldmath$\mu$}} \in \Lambda} \sum_{(nj) \in \cL} \mu_{nj} + \sum_{n,d} q_{nd} \bigl(
 \sum_{f\in \cF}x_f I_{\{n = b(f),e(f)=d\}} \nonumber\\
 & & \hspace{10mm} + \sum_{(ln)\in \cL} \mu^d_{ln} - \sum_{(nj)\in \cL} \mu^d_{nj} \bigr) \\
 &=&\min_{\mbox{{\boldmath$\mu$}} \in \Lambda} \Bigl( - \sum_{(nj) \in \cL} \sum_{d} \mu_{nj}^d \bigl( q_{nd}-q_{jd}-1 \bigr) \nonumber \\
& & \hspace{10mm} - \sum_{n,d} q_{nd} \sum_{f\in \cF}x_f I_{\{n = b(f),e(f)=d\}} \Bigr). \nonumber
\end{eqnarray}
If the Lagrange multipliers are known, then the optimal {\boldmath$\mu$} can be found by solving
\begin{eqnarray*}
\max_{\mbox{{\boldmath$\mu$}}\in \Lambda} \sum_{(nj) \in \cL} \mu_{nj} w_{nj}
\end{eqnarray*}
where $w_{nj} = \displaystyle \max_{d} (q_{nd}-q_{jd}-1).$ The form of the constraints in (\ref{eqn:minhop_opt}) suggests the following update algorithm to compute $q_{nd}:$
\begin{eqnarray} \label{eqn:lag_update}
q_{nd} [t+1] = \Bigl[ q_{nd}[t] + \frac{1}{M} \bigl( \sum_{f\in \cF}x_f I_{\{n = b(f),e(f)=d\}} \nonumber \\
+ \sum_{(ln)\in \cL} \mu^d_{ln} - \sum_{(nj)\in \cL} \mu^d_{nj} \bigr) \Bigr]^{+}
\end{eqnarray}
where  $\frac{1}{M}$ is a step-size parameter. Notice that $Mq_{nd}[t]$ looks very much like a queue update equation, except for the fact that arrivals into $Q_{nd}$ from other links may be smaller than $\mu_{ln}^d$ when $Q_{ld}$ does not have enough packets. This suggests the following algorithm.

\noindent\textbf{Min-resource routing by back-pressure:} At time
slot $t,$
\begin{itemize}

\item Each node $n$ maintains a separate queue of packets
for each destination $d$; its length is denoted $Q_{nd}[t]$.
Each link is assigned a weight
% equal to the maximum differential backlog
%    minus a constant $\gamma :$
\begin{equation}
w_{nj}[t] = \max_{d}
\left( \frac{1}{M}Q_{nd}[t] - \frac{1}{M}Q_{jd}[t] - 1
\right), \label{eqn:minhop_diff1}
\end{equation}
where $M>0$ is a parameter.

\item Scheduling/routing rule:
\begin{equation}
\pi^*[t] \in \arg\max_{\pi \in \Gamma} \sum_{(nj)\in \pi} c_{nj} w_{nj}[t].
\label{eqn:minhop_sched}
\end{equation}

\item For each activated link $(nj) \in \pi^*[t]$ we remove $c_{nj}$ packets from $Q_{nd_{nj}^*[t]}$ if possible, and transmit those packets to $Q_{jd_{nj}^*[t]},$ where $d_{nj}^*[t]$ achieves the maximum in (\ref{eqn:minhop_diff1}).

\end{itemize}

Note that the above algorithm does not change if we replace the
weights in (\ref{eqn:minhop_diff1}) by the following, re-scaled
ones:
\begin{equation}
w_{nj}[t] = \max_{d}
\left( Q_{nd}[t] - Q_{jd}[t] - M
\right), \label{eqn:minhop_diff}
\end{equation}
and therefore, compared with the traditional back-pressure
scheduling/routing, the only difference is that each link
weight is equal to the maximum differential backlog {\em minus
parameter $M$}. ($M=0$ reverts the algorithm to the traditional one.)
For simplicity, we call this algorithm {\it$M$-back-pressure algorithm}.

The performance of the stationary process which is ``produced''
by the algorithm with fixed parameter $M$ is within $o(1)$ of
the optimal as $M$ goes to $\infty$ (analogous to the proofs in \cite{neemodli05, sto05}; see also the related proof in \cite{erysri05, erysri06}):
$$
\left| ~\bE \left[\sum_{(nj) \in \cL} \mu_{nj}[\infty]\right] - \sum_{(nj)\in \cL} \mu_{nj}^*
~\right|= o(1),
$$
where $\mu^*$ is an optimal solution to (\ref{eqn:minhop_opt}).

Although $M$-back-pressure algorithm could reduce the delay by forcing flows to go through shorter routes, simulations indicate a significant problem with the basic algorithm presented above. A link can be scheduled only if the back-pressure of at least one destination is greater than or equal to $M.$ Thus, at light to moderate traffic loads, the delays could be high since the back-pressure may not build up sufficiently fast.
In order to overcome all these adverse issues, we develop a new routing algorithm in the following section. The solution also simplifies the queueing data structure to be maintained at each node.

\section{PARN: Packet-by-Packet Adaptive Routing and Scheduling Algorithm For Networks}
\label{sec:adaptive-routing}

In this section, we present our adaptive routing and scheduling algorithm. We will call it PARN (Packet-by-Packet Adaptive Routing for Networks) for ease for repeated reference later. First, we introduce the queue structure that is used in PARN.

In the traditional back-pressure algorithm, each node $n$ has to maintain a queue $q_{nd}$ for each destination $d.$ Let $|\cN|$ and $|\cD|$ denote the number of nodes and the number of destinations in the network, respectively. Each node maintains $|\cD|$ queues. Generally, each pair of nodes can communicate along a path connecting them. Thus, the number of queues maintained at each node can be as high as one less than the number of nodes in the network, i.e., $|\cD|$=$|\cN|-1.$

Instead of keeping a queue for every destination, each node $n$ maintains a queue $q_{nj}$ for every neighbor $j,$ which is called a {\it real queue}. Notice that real queues are per-neighbor queues. Let $J_n$ denote the number of neighbors of node $n,$ and let $J_{max} = \max_{n} J_n.$ The number of queues at each node is no greater than $J_{max}.$ Generally, $J_{max}$ is much smaller than $|\cN|.$ Thus, the number of queues at each node is much smaller compared with the case using the traditional back-pressure algorithm.

In additional to real queues, each node $n$ also maintains a counter, which is called {\it shadow queue,} $p_{nd}$ for each destination $d.$ Unlike the real queues, counters are much easier to maintain even if the number of counters at each node grows linearly with the size of the network. A back-pressure algorithm run on the shadow queues is used to decide which links to activate. The statistics of the link activation are further used to route packets to the per-next-hop neighbor queues mentioned earlier. The details are explained next.

\subsection{Shadow Queue Algorithm -- $M$-back-pressure Algorithm}

The shadow queues are updated based on the movement of fictitious entities called shadow packets in the network. The movement of the fictitious packets can be thought of as an exchange of control messages for the purposes of routing and schedule. Just like real packets, shadow packets arrive from outside the network and eventually exit the network. The external shadow packet arrivals are general as follows:
%
%To run the shadow queue algorithm, we should generate several exogenous incentive shadow traffic flows for the network. Obviously, the shadow traffic flows should be related to the real traffic flows. The following method shows how we generate the incoming shadow traffic:
when an exogenous packet arrives at node $n$ to the destination $d,$ the shadow queue $p_{nd}$ is incremented by $1,$ and is further incremented by $1$ with probability $\varepsilon$ in addition. Thus, if the arrival rate of a flow $f$ is $x_{f},$ then the flow generates ``shadow traffic'' at a rate $x_f(1+\varepsilon).$ In words, the incoming shadow traffic in the network is $(1+\varepsilon)$ times of the incoming real traffic.

The back-pressure for destination $d$ on link $(nj)$ is taken to be $$w_{nj}^d[t]=p_{nd}[t]-p_{jd}[t]-M,$$ where $M$ is a properly chosen parameter. The choice of $M$ will be discussed in the simulations section.

The evolution of the shadow queue $p_{nd}[t]$ is
\begin{eqnarray}\label{eq: shadow queue dynamics}
\begin{array}{lll} p_{nd}[t+1] &=&p_{nd}[t]
- \displaystyle \sum_{j: (nj) \in \cL} I_{\{d^*_{nj}[t]=d\}}\,\hat{\mu}_{nj}[t]
\\&&+ \displaystyle \sum_{l: (ln) \in \cL} I_{\{d^*_{ln}[t]=d\}} \,\hat{\mu}_{ln}[t]
\\&&+\displaystyle\sum_{f \in \cF} I_{\{b(f)=n,e(f)=d\}}\, \hat a_f[t],
\end{array}
\end{eqnarray}
where $\hat{\mu}_{nj}[t]$ is the number of shadow packets transmitted over
link $(nj)$ in time slot $t$, $d^*_{nj}[t]$ is the destination that has the maximum weight on link $(nj),$ and $\hat a_f[t]$ is the number of shadow packets generated by flow $f$ at time $t.$ The number of shadow packets scheduled over the links at each time instant is determined by the back-pressure algorithm in equation (\ref{eqn:minhop_sched}).

From the above description, it should be clear that the shadow algorithm is the same as the traditional back-pressure algorithm, except that it operates on the shadow queueing system with an arrival rate slightly larger than the real external arrival rate of packets.
Note the shadow queues do not involve any queueing data structure at each node; there are no packets to maintain in a FIFO order in each queue. The shadow queue is simply a counter which is incremented by $1$ upon a shadow packet arrival and decremented by $1$ upon a departure.

The back-pressure algorithm run on the shadow queues is used to activate the links. In other words, if $\pi^*_{nj}=1$
in (\ref{eqn:minhop_sched}), then link $(nj)$ is activated and packets are served from the real queue at the link in a first-in, first-out fashion. This is, of course, very different from the traditional back-pressure algorithm where a link is activated to serve packets to a particular destination. Thus, we have to develop a routing scheme that assigns packets arriving to a node to a particular next-hop neighbor so that the system remains stable. We design such an algorithm next.

%Note that shadow queues are not real queues, so no shadow packets are really transmitted over any link. The counters (shadow queues) are simply updated as determined by the shadow queue algorithm. In wireline networks, the shadow queue algorithm can be deployed in a distributed manner by occasional exchanges of messages between neighboring nodes. In wireless networks, the back-pressure algorithm needs a central controller because of the existence of the conflicting links. However we can resort to LWF algorithm instead of the $M$-back-pressure algorithm to reduce the complexity by paying the price of the reduction of the network capacity. We can also resort to the class of queue-length based Carrier Sense Multiple Access (CSMA) algorithms to deploy PARN in a distributed manner.

%By running shadow queue algorithm, we acquire the knowledge of the shadow traffic on each link at each time slot . Based on this knowledge, we come out our adaptive routing algorithms.

\subsection{Adaptive Routing Algorithms}

Now we discuss how a packet is routed once it arrives at a node.
%
%\if 0
%Recall that the real queues are per-neighbor queues and the number of queues at a node is equal to the number of its neighbors unlike the traditional back-pressure algorithm which requires per-destination queues. As in \cite{buisristo09}, such reduction in queueing complexity is obtained by maintaining per-flow counters which are much easier to update than queues. However, unlike \cite{buisristo09}, we still have to determine how packets are to be routed in the network.
%\fi
%
%We explore two algorithms that are competent for adaptive routing.
%
%\subsubsection{Probabilistic Splitting Algorithm}
%Once a packet arrives at node $n,$ we should decide the next hop to which it is routed. One method is to decide the next hop based on a probabilistic routing table. The probabilistic routing table is time-varing and updated based on the outgoing shadow traffic information at each node at every slot. Of course, the most recent shadow traffic information is more important because it captures the timely information.
%Next, we will talk about how we choose the probability distribution at each node $n.$ The idea is that we want the real traffic behaves like the shadow traffic, and since the link activation process can stabilize the shadow traffic, it should stabilize the real traffic too.
%
Let us define a variable $\sigma_{nj}^d[t]$ to be the number of shadow packets ``transferred" from node $n$ to node $j$ for destination $d$ during time slot $t$ by the shadow queue algorithm.
Let us denote by $\bar{\sigma}_{nj}^d$ the expected value of $\sigma_{nj}^d[t]$, when
the shadow queueing process is in a stationary regime; let  $\hat{\sigma}_{nj}^d[t]$
denote an estimate of $\bar{\sigma}_{nj}^d$, calculated at time $t$. (In the simulations
we use the exponential averaging, as specified in the next section.)

\begin{figure}[h]
    \begin{center}
        \includegraphics[width=50mm]{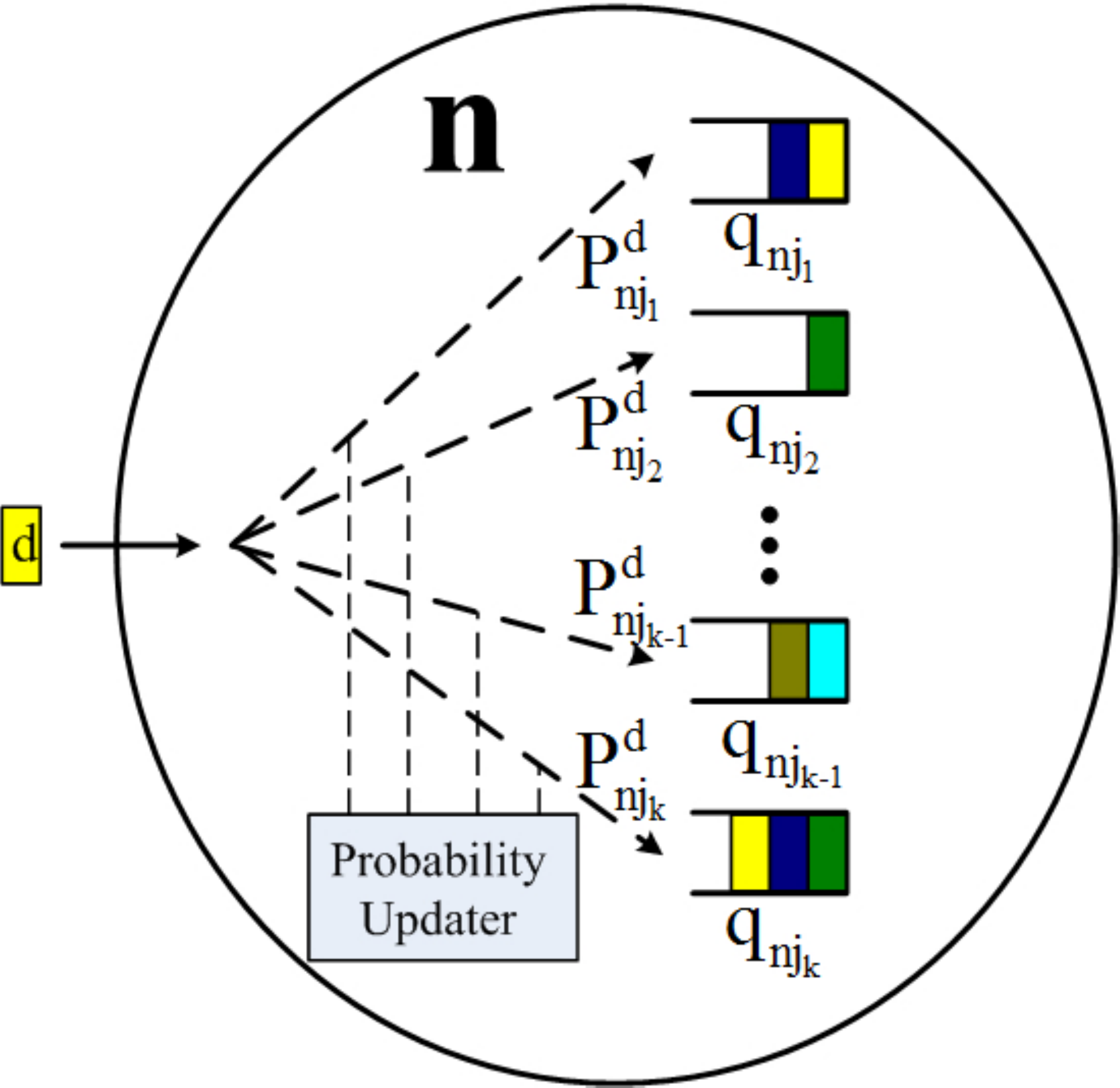}
    \end{center}
        \caption{Probabilistic splitting algorithm in Node $n$.}
        \label{fig:probability_splitting}
\end{figure}

At each time slot, the following sequence of operations occurs at
each node $n.$ A packet arriving at node $n$ for destination $d$ is
inserted in the real queue $q_{nj}$ for next-hop neighbor $j$ with
probability
\begin{eqnarray}\label{eq: prob split}
P_{nj}^d[t]=\displaystyle \frac{\hat{\sigma}_{nj}^d[t]}
{\sum_{k: (nk)\in \cL} \hat{\sigma}_{nk}^d[t]}.
\end{eqnarray}
Thus, the estimates $\hat{\sigma}_{nj}^d[t]$ are used to perform routing operations: in
today's routers, based on the destination of a packet, a packet is
routed to its next hop based on routing table entries. Instead, here, the $\bar{\sigma}$'s
are used to probabilistically choose the next hop for a packet. Packets waiting at link $(nj)$ are transmitted over the link when that link is scheduled (See Figure~\ref{fig:probability_splitting}).

The first question that one must ask about the above algorithm is
whether it is stable if the packet arrival rates from flows are
within the capacity region of the multi-hop network. This is a difficult question, in general.
Since the shadow queues are positive recurrent,
``good'' estimates $\hat{\sigma}_{nj}^d[t]$ can be maintained by simple averaging
(e.g. as specified in the next section), and therefore
the probabilities in (\ref{eq: prob split})
will stay close to their ``ideal'' values
$$
\bar{P}_{nj}^d=\displaystyle \frac{\bar{\sigma}_{nj}^d}
{\sum_{k: (nk)\in \cL} \bar{\sigma}_{nk}^d}.
$$
The following theorem asserts that the real queues are stable if $P_{nj}^d$
are fixed at $\bar P_{nj}^d.$

\theorem
\label{real-stability}
Suppose, $P_{nj}^d[t]\equiv \bar P_{nj}^d$.
Assume that there exists a delta such that $\{x_f(1+\epsilon +\delta)\}$ lies in $\Gamma$. Let $a_f[t]$ be the number of packets arriving from flow $f$ at time slot $t,$ with $E(a_f[t])=x_f$ and $E(a_f[t]) <\infty.$ Assume that the arrival process  is independent across time slots and flows (this assumption can be considerably relaxed). Then, the Markov chain,
jointly describing the evolution of shadow queues and real FIFO queues (whose state
include the destination of the real packet in each position of each FIFO queue),
is positive recurrent.

\begin{IEEEproof}
The key ideas behind the proof are outlined. The details are similar to the proof
in \cite{buisristo09a} and are omitted.

\begin{itemize}

\item
The average rate at which packets arrive to link $(nj)$ is strictly smaller than the capacity allocated to the link by the shadow process if $\varepsilon > 0$. (This fact is verified in Appendix~\ref{sec:stability-proof}.)

\item
It follows that the fluid limit of the real-queue process is same as that
of the networks in \cite{bra96}. Such fluid limit is stable \cite{bra96},
which implies the stability of our process as well.
\end{itemize}

\end{IEEEproof}

\section{Implementation Details}
\label{sec:implement_detail}

The algorithm presented in the previous section ensures that the queue lengths are stable. In this section, we discuss a number of enhancements to the basic algorithm to improve performance.

\subsection{Exponential Averaging}
To compute $\hat{\sigma}_{nj}^d[t]$  we use the following
iterative
exponential averaging algorithm:
\begin{eqnarray}\label{eq: averaging_new}
\hat{\sigma}_{nj}^d[t]=(1-\beta) \; \hat{\sigma}_{nj}^d[t-1] +
\beta\;{\sigma}_{nj}^d[t],
\end{eqnarray}
where $0 < \beta < 1.$

\subsection{Token Bucket Algorithm}

Computing the average shadow rate $\hat \sigma_{nj}^d[t]$ and generating random numbers for routing packets may impose a computational overhead of routers which should be avoided if possible. Thus, as an alternative, we suggest the following simple algorithm.
At each node $n,$ for each next-hop neighbor $j$ and each
destination $d,$ maintain a token bucket $r_{nj}^d.$
Consider the shadow traffic as a guidance of the real traffic, with
 tokens {\em removed}
as shadow packets traverse the link. In detail,
the token bucket is decremented by $\sigma_{nj}^d [t]$ in each time slot, but
cannot go below the lower bound $0$:
$$
r_{nj}^d[t] = \max\{r_{nj}^d[t-1] - \sigma_{nj}^d [t], 0\}.
$$
When $r_{nj}^d[t-1] - \sigma_{nj}^d [t]<0$, we say that
$\sigma_{nj}^d [t]-r_{nj}^d[t-1]$ tokens (associated with bucket
$r_{nj}^d$) are ``wasted'' in slot $t$.
Upon a packet arrival at
node $n$ for destination $d,$ find the token bucket $r_{n j^*}^d$
which has the smallest number of tokens (the minimization is over
next-hop neighbors $j$), breaking ties arbitrarily, add the packet
to the corresponding real queue $q_{nj^*}$ and add one token to the
corresponding bucket:
\begin{equation}
\label{eq-up}
%\label{eq-up-clip}
r_{nj^*}^d[t] = r_{nj^*}^d[t-1] + 1.
\end{equation}

To explain how this algorithm works, denote by $\bar{\sigma}_{nj}^d$
the average value of $\sigma_{nj}^d[t]$ (in stationary regime), and
by $\eta_{n}^d$ the average rate at which real packets for
destination $d$ arrive at node $n$. Due to the fact that real
traffic is injected by each source at the rate strictly less than
the shadow traffic, we have
\begin{equation}
\label{eq-subcritical} \eta_{n}^d < \sum_j
\bar{\sigma}_{nj}^d.
\end{equation}
For a single-node network, (\ref{eq-subcritical}) just means that arrival rate is less than available capacity. More generally, it is an assumption that needs to be proved. However, here our goal is to provide an intuition behind the token bucket algorithm, so we simply assume (\ref{eq-subcritical}).
Condition (\ref{eq-subcritical}) guarantees
that the token processes are {\em stable} (that is, roughly, they
cannot runaway to infinity) since the total arrival rate
to the token buckets at a node is less than the total service rate
and the arrivals employ a join-the-shortest-queue discipline.
Moreover, since $r_{nj}^d[t]$ are random
processes, the token buckets will ``hit 0''
in a non-zero fraction of time slots, except in some degenerate
cases; this in turn means that the arrival rate of packets
at the token bucket must be less than the token generation rate:
\begin{equation}
\label{eq-subcritical2} \eta_{nj}^d < \bar{\sigma}_{nj}^d,
\end{equation}
where $\eta_{nj}^d$ is the actual rate at which packets
arriving at $n$ and destined for $d$ are routed along link $(nj)$.
Inequality (\ref{eq-subcritical2}) thus describes the idea of the
algorithm.

Ideally, in addition to (\ref{eq-subcritical2}), we would like to
have the ratios $\eta_{nj}^d / \bar{\sigma}_{nj}^d$ to be
equal across all $j$, i.e., the real packet arrival rates at the outgoing links of
a node should be proportional to the shadow service rates. It is not difficult to see that if $\varepsilon$ is very small, the proportion will be close to ideal. In general,
the token-based algorithm does not guarantee
that, that is why it is an approximation.

Also, to ensure implementation correctness, instead of
(\ref{eq-up}), we use
\begin{equation}
\label{eq-up-clip} r_{nj_*}^d[t] = \min\{r_{nj_*}^d[t-1] + 1, B\},
\end{equation}
i.e., the value of $r_{nj_*}^d[t]$ is not allowed to go above some
relatively large value $B$, which is a parameter of the order of
$O(1/\epsilon)$. Under ``normal circumstances'', $r_{nj_*}^d[t]$
``hitting'' ceiling $B$ is a rare event, occurring due to the process
randomness. The main purpose of having the upper bound $B$ is to
detect serious anomalies when, for whatever reason, the condition
(\ref{eq-subcritical}) ``breaks'' for prolonged periods of time --
such situation is detected when any $r_{nj_*}^d[t]$ hits the upper bound
$B$ frequently.

%\subsection{The Role of $M$}
%The parameter $M$ in the modified back-pressure algorithm should be chosen properly in simulation. The value $M$ affects the the average packet or file delay in the network. If $M$ is chosen to be zero, we run the traditional back-pressure algorithm on shadow queues. As we know, the back-pressure algorithm results in large delay especially in light traffic and out-of-sequence delivery problem in TCP. By choosing $M$ large,  we minimize the sum of resources $\sum_{(nj)\in \cL}\mu_{nj}$ in the network\cite{buisristo09}. In other words, short paths of each flow can be discovered automatically by choosing $M$ properly. On the other side,  choosing too large $M$ will result in the slow convergence to the optimal point and the queue backlogs at each node. Furthermore, last-packet problem will be a disaster if we don't have further action when we choose large $M.$

\subsection{Extra Link Activation}

Under the shadow back-pressure algorithm, only links with back-pressure greater than or equal to $M$ can be activated. The stability theory ensures that this is sufficient to render the real queues. On the other hand, the delay performance can still be unacceptable. Recall that the parameter M was introduced to discourage the use of unnecessarily long paths. However, under light and moderate traffic loads, the shadow back-pressure at a link may be frequently less than $M$, and thus, packets at such links may have to wait a long time before they are processed. One way to remedy the situation is to activate additional links beyond those activated by the shadow back-pressure algorithm.

The basic idea is as follows: in each time slot,
first run the shadow back-pressure algorithm. Then, add additional links to make the
schedule maximal. If the extra activation procedure depends only on the state
of shadow queues (but beyond that, can be random and/or arbitrarily complex),
then the stability result of Theorem~\ref{real-stability} still holds
(with essentially same proof).
Informally, the stability prevails, because the shadow algorithm alone
provides sufficient average throughput on each link, and adding extra capacity
``does not hurt''; thus, with such extra activation, a certain degree
of ``decoupling'' between routing (totally controlled by shadow queues) and scheduling
(also controlled by shadow queues, but not completely) is achieved.

For example, in the case of wireline networks,
by the above arguments, all links can be activated all the time.
The shadow routing algorithm ensures that the arrival rate at each link is less than its capacity.
In this case the {\em complete} decoupling of routing and scheduling occurs.

In practice, activating extra links which have large queue backlogs leads to better performance than activating an arbitrary set of extra links.
However, in this case, the extra activation procedure depends on the state of
real queues which makes
the issue of validity of an analog of Theorem~\ref{real-stability} much more subtle. We believe that the argument in this subsection provides
a good motivation for our algorithm, which is confirmed by simulations.

\subsection{The Choice of the Parameter $\varepsilon$}

From basic queueing theory, we expect the delay at each link to be inversely proportional to the mean capacity minus the arrival rate at the link. In a wireless network, the capacity at a link is determined by the shadow scheduling algorithm. This capacity is guaranteed to be at least equal to the shadow arrival rate. The arrival rate of real packets is of course smaller. Thus, the difference between the link capacity and arrival rate could be proportional to epsilon. Thus, epsilon should be sufficiently large to ensure small delays while it should be sufficiently small to ensure that the capacity region is not diminished significantly. In our simulations, we found that choosing $\varepsilon=0.1$ provides a good tradeoff between delay and network throughput.

In the case of wireline networks, recall from the previous subsection that all links are activated. Therefore, the parameter epsilon plays no role here.

\section{Extension to The Network Coding Case}
\label{sec:extend-network-coding}
%In the previous sections, we don't fully use the nature of the wireless channel. Due to the sharing of the wireless medium, a set of links can be activated with no fear of collisions. The capacity region is expanded by broadcast transmission with network coding. Our main objective in this section is to develop our mechanism with the presence of the broadcast transmission which can support larger throughput without sacrificing much on the delay.

\begin{figure}[h]
\begin{center}
    \includegraphics[scale=0.5]{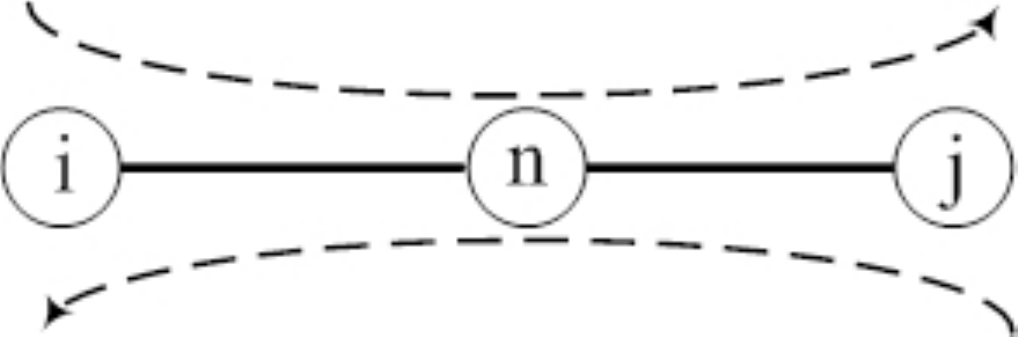}
    \caption{Network coding opportunity.}
    \label{coding_opp}
\end{center}
\end{figure}

In this section, we extend our approach to consider networks where network coding is used to improve throughput. We consider a simple form of network coding illustrated in Figure~\ref{coding_opp}. When $i$ and $j$ each have a packet to send to the other through an intermediate relay $n$, traditional transmission requires the following set of transmissions: send a packet a from $i$ to $n$, then $n$ to $j$, followed by $j$ to $n$ and $n$ to $i$. Instead, using network coding, one can first send from $i$ to $n$, then $j$ to $n$, XOR the two packets and broadcast the XORed packet from $n$ to both $i$ and $j$. This form of network coding reduces the number of transmissions from four to three.  However, the network coding can only improve throughput only if such coding opportunities are available in the network. Routing plays an important role in determining whether such opportunities exist. In this section, we design an algorithm to automatically find the right tradeoff between using possibly long routes to provide network coding opportunities and the delay incurred by using long routes.

%The network coding can only be used when there is coding opportunity. In the paper we don't consider overhearing which induces other complexity issues not only in data structure but also in algorithm design, and we only consider the case that the number of the intended receivers of a broadcast is 2. Therefore, the coding opportunity happens only when two flows overlap with two consecutive links from opposite directions. See the code opportunity example in Figure~\ref{coding_opp}. We also assume that the network coding scheme used in our network model is myopia such that the coded packets are decoded just after received at the next-hop node. Because the routes are not fixed under our scheme, we can create coding opportunities by routing packets to the particular next-hop neighbors. Our adaptive routing algorithm will automatically help us create the coding opportunity if it is necessary.

\subsection{System Model}
We still consider the wireless network represented by the graph $\cG=(\cN,\cL).$ Let $x_f$ be the rate (packets/slot) at which packets are generated by flow $f.$ To facilitate network coding, each node must not only keep track of the destination of the packet, but also remember the node from which a packet was received. Let $\mu_{lnj}^d$ be the rate at which packets received from either node $l$ or flow $l$, destined for node $d$, are scheduled over link $(nj)$.
Note that, for compactness of notation, we allow $l$ in the definition of $\mu_{lnj}^d$ to denote either a flow or a node. We assume $\mu_{lnj}^d$ is zero when such a transmission is not feasible, i.e., when $n$ is not the source node or $d$ is not the destination node of flow $l$,  or if $(ln)$ or $(nj)$ is not in $\cL$.
At node $n,$ the network coding scheme may generate a coded packet by ``XORing" two packets received from previous-hop nodes $l$ and $j$ destined for the destination nodes $d$ and $d'$ respectively, and broadcast the coded packet to nodes $j$ and $l.$ Let $\mu_{n|jl}^{d,d'}$ denote the rate at which coded packets can be transferred from node $n$ to nodes $j$ and $l$ destined for nodes $d$ and $d',$ respectively. Notice that, due to symmetry, the following equality holds $\mu_{n|jl}^{d,d'} = \mu_{n|lj}^{d',d}.$ Assume $\mu_{n|jl}^{d,d'}$ to be zero if at least one of $(nl),(ln),(nj)$ and $(jn)$ doesn't belong to $\cL.$ Note that $\mu_{lnj}^d=0$ when $d=l$ or $d=n,$ and $\mu_{n|jl}^{d,d'}=0$ when $d=n$ or $d'=n.$

There are two kinds of transmissions in our network model: point-to-point transmissions and broadcast transmissions. The total point-to-point rate at which packets received externally or from a previous-hop node are scheduled on link $(nj)$ and destined to $d$ is denoted by
\begin{eqnarray*}
\mu_{nj,pp}^{d} = \sum_{l:l \in \cF} \mu_{lnj}^d + \sum_{l:l \in \cN} \mu_{lnj}^d,
\end{eqnarray*}
and the total broadcast rate at which packets scheduled on  link $(nj)$ destined to $d$ is denoted by
\begin{eqnarray*}
\mu_{nj,broad}^d = \sum_{d'} \sum_{l: l \neq j} \mu_{n|jl}^{d,d'}.
\end{eqnarray*}
The total point-to-point rate on link $(nj)$ is denoted by
\begin{eqnarray*}
\mu_{nj,pp} = \sum_d \mu_{nj,pp}^{d}
\end{eqnarray*}
and the total broadcast rate at which packets are broadcast from node $n$ to nodes $j$ and $l$ is denoted by
\begin{eqnarray*}
\mu_{n|jl} = \sum_{d'} \sum_{d} \mu_{n|jl}^{d,d'}.
\end{eqnarray*}
Let {\boldmath$\mu$} be the set of rates including all point-to-point transmissions and broadcast transmissions, i.e.,
\begin{eqnarray*}
\mbox{{\boldmath$\mu$}} &=& \{\{\mu_{nj,pp}\}_{(nj)}, \{\mu_{n \mid jl}\}_{(n \mid jl)}\}.
\end{eqnarray*}

The multi-hop traffic should also satisfy the flow conservation constraints.\\
\noindent{\it Flow conservation constraints:} For each node $n, $ each neighbor $j,$ and each destination $d,$ we have
\begin{eqnarray}
\mu_{nj,pp}^d + \mu_{nj,broad}^{d} \le \sum_{k} \mu_{njk}^d + \sum_{d'} \sum_{k: k \neq n}\mu_{j|kn}^{d,d'},
\label{bro_constraint1}
\end{eqnarray}
where the left-hand side denotes the total incoming traffic rate at link $nj$ destined to $d,$ and the right-hand side denotes the total outgoing traffic rate from link $nj$ destined to $d.$ For each node $n$ and each destination $d,$ we have
\begin{eqnarray}
\sum_{f \in \cF}x_f I_{\{b(f)=n, e(f)=d\}} \le \sum_{f \in \cF} \sum_{j \in \cN} \mu_{fnj}^d,
\label{bro_constraint2}
\end{eqnarray}
where $I$ denotes the indicator function.

\subsection{Links and Schedules}
We allow broadcast transmission in our network model. In order to define a schedule, we first define two kinds of ``links:" the point-to-point link and the broadcast link. A point-to-point link $(nj)$ is a link that supports point-to-point transmission, where $(nj) \in \cL;$  A broadcast link $(n|lj)$ is a ``link" which contains links $(nl)$ and $(nj)$ and supports broadcast transmission. Let $\cB$ denote the set of all broadcast links, thus $(n|lj) \in \cB.$ Let $\bar \cL$ be the union of the set of the point-to-point links $\cL$ and the set of the broadcast links $\cB,$ i.e., $\bar \cL = \cL \cup \cB.$

We let $\Gamma'$ denote the set of links that can be activated simultaneously. By abusing notation, $\Gamma'$ can be thought of as a set of vectors where each vector is a list of $1$'s or $0$'s where a $1$ corresponds to an active link and a $0$ corresponds to an inactive link.
Then, the capacity region of the network for $1$-hop traffic is the convex hull of all schedules, i.e., $\Lambda' = co(\Gamma').$  Thus, $\mu \in \Lambda'$.

\subsection{Queue Structure and Shadow Queue Algorithm}
Each node $n$ maintains a set of counters, which are called {\it shadow queues,} $p_{lnd}$ for each previous hop $l$ and each destination $d,$ and $p_{0nd}$ for  external flows destined for $d$ at node $n.$ Each node $n$ also maintains a real queue, denoted by $q_{lnj},$ for each previous hop $l$ and each next-hop neighbor $j,$ and $q_{0nj}$ for external flows with their next hop $j.$

By solving the optimization problem with flow conservation constraints, we can work out the back-pressure algorithm for network coding case (see the brief description in Appendix \ref{sec:find_backpressure_networkcoding}). More specifically, for each link $(nj) \in L$ in the network and for each destination $d,$ define the {\it back-pressure} at every slot to be
\begin{eqnarray}\label{eq: Wireless point-to-point weight}
\begin{array}{l} w_{nj}^d[t] = \displaystyle \max_{l: (ln) \in \cL
\mbox{\footnotesize{ or }} l=0} w_{lnj}^d[t]  \\
\mbox{where }
w_{lnj}^d[t]  =  p_{lnd}[t]  -p_{njd}[t]-M ,\\
\mbox{and } l^*_{nj}[t] = \displaystyle  \arg\max_{l: (ln) \in \cL
\mbox{\footnotesize{ or }} l=0} w_{lnj}^d[t].
\end{array}
\end{eqnarray}
For each broadcast at node $n$ to nodes $j$ and $l$ destined for $d$ and $d',$ respectively, define the {\it back-pressure} at every slot to be
\begin{eqnarray}\label{eq: Wireless brodcast weight}
w_{n|jl}^{d,d'}[t] = w_{lnj}^d[t] + w_{jnl}^{d'}[t].
\end{eqnarray}
The weights associated with each point-to-point link $(nj) \in \cL$ and each broadcast link $(n|jl)$ are defined as follows
\begin{eqnarray}\label{eq: Wireless rates}
\begin{array}{l}
w_{nj}[t] = \displaystyle  \max_d \{ w_{nj}^d[t] \},\\
w_{n \mid jl}[t] = \displaystyle \max_{d,d'} \{ w_{n \mid jl}^{d,d'}[t] \},\\
\mbox{with } d^*_{nj}[t]  = \displaystyle  \arg \max_d \{ w_{nj}^d[t] \},\\
\{d, d'\}^*_{n \mid jl}[t] = \displaystyle  \arg \max_{d,d'} \{ w_{n
\mid jl}^{d,d'}[t] \}.
\end{array}
\end{eqnarray}
The rate vector {\boldmath$\tilde \mu^* $}$[t]$ at each time slot is chosen to satisfy
\begin{eqnarray*}\label{eq: Wireless weight rates}
\begin{array}{rl}
\mbox{\boldmath$\tilde\mu^*$}[t] \in \arg \displaystyle
\max_{\mbox{\boldmath$\tilde{\mu}$} \in \Gamma'} \Big\{
\displaystyle \sum_{(nj) \in \cL} \tilde{\mu}_{nj,pp} w_{nj}[t] \\
 + \displaystyle \sum_{(n \mid jl) \in \cB} \tilde{\mu}_{n \mid jl}
w_{n \mid jl}[t]\Big\}.
\end{array} \end{eqnarray*}
By running the shadow queue algorithm in network coding case, we get a set of activated links in $\bar \cL$ at each slot.

Next we describe the evolution of the shadow queue lengths in the network. Notice that the shadow queues at each node $n$ are distinguished by their previous hop $l$ and their destination $d,$ so $p_{lnd}$ only accepts the packets from previous hop $l$ with destination $d.$ The similar rule should be followed when packets are drained from the shadow queue $p_{lnd}.$ We assume the departures occur before arrivals at each slot, and the evolution of queues is given by
\begin{eqnarray}\label{eq: shadow queue dynamics2}
p_{lnd}[t+1] &=& \Bigl[p_{lnd}[t] - \displaystyle \sum_{j \in \cN} \tilde \mu_{nj,pp}^*[t] I_{\{l=l_{nj}^*, d=d_{nj}^*\}}\nonumber \\
&-&  \displaystyle{\sum_{d' \in \cN} \sum_{j\in \cN}} \tilde\mu_{n|jl}^*[t] I_{\{\{d,d'\}= \{d,d'\}_{n|jl}^*  \}} \Bigr]^+\nonumber \\
&+& \displaystyle \sum_{k \in \cN} \hat\mu_{kln}^d[t] I_{\{k=l_{ln}^*, d=d_{ln}^*\}} \\
&+& \displaystyle \sum_{k \in \cN} \sum_{d' \in \cN} \hat \mu_{l | nk}^{d,d'}[t] I_{\{\{d,d'\}= \{d,d'\}_{l | nk}^*  \}}\nonumber \\
&+& \displaystyle \sum_{f \in \cF} \hat a_f[t] I_{\{b(f)=n, e(f)=d, l=0\}}, \nonumber
\end{eqnarray}
where $\hat\mu_{kln}^d[t]$ is the actual number of shadow packets scheduled over link $(ln)$ and destined for $d$ from the shadow queue $p_{kld}$ at slot $t,$ $\hat \mu_{l | nk}^{d,d'}[t] $ is the actual number of coded shadow packets transfered from node $l$ to nodes $n$ and $k$ destined for nodes $d$ and $d'$ at slot $t,$ and $\hat a_f$ denotes the actual number of shadow packets from external flow $f$ received at node $n$ destined for $d.$

\subsection{Implementation Details}
The implementation details of the joint adaptive routing and coding algorithm are similar to the case with adaptive routing only, but the notation is more cumbersome. We briefly describe it here.

\subsubsection{Probabilistic Splitting Algorithm}
The probabilistic splitting algorithm chooses the next hop of the packet based on the probabilistic routing table. Let $P_{lnj}^d[t]$ be the probability of choosing node $j$ as the next hop once a packet destined for $d$ receives at node $n$ from previous hop $l$ or from external flows, i.e., $l=0$ at slot $t.$ Assume that $P_{lnj}^d[t]=0$ if $(nj) \not \in \cL.$ Obviously, $\sum_{j \in \cN} P_{lnj}^d[t]=1.$
Let $\sigma_{lnj}^d[t]$ denote the number of potential shadow packets ``transferred" from node $n$ to node $j$ destined for $d$ whose previous hop is $l$ during time slot $t.$ Notice that the packet comes from an external flow if $l=0.$
Also notice that $\sigma_{lnj}^d[t]$ is contributed by shadow traffic point-to-point transmission as well as shadow traffic broadcast transmission, i.e.,
\begin{eqnarray*}
\sigma_{lnj}^d[t] \ \ = & \mu_{nj,pp}^*[t] I_{\{ l=l_{nj}^*[t] , d= d_{nj}^*[t] \}} \\
&+ \displaystyle \sum_{d' \in \cN} \mu_{n|jl}^*[t] I_{\{ \{d,d'\}=\{d,d'\}_{n|jl}^*[t] \}}.
\end{eqnarray*}
We keep track of the the average value of $\sigma_{lnj}^d[t]$ across time by using the following updating process:
\begin{eqnarray}\label{eq:update_average_sigma_netcode}
\hat\sigma_{lnj}^d[t]=(1-\beta) \hat\sigma_{lnj}^d[t-1]+\beta \sigma_{lnj}^d[t],
\end{eqnarray}
where $0 \le \beta \le 1.$
The splitting probability $P_{lnj}^d[t]$ is expressed as follows:
\begin{eqnarray}\label{eq:splitting_probability_netcode}
P_{lnj}^d[t] = {\hat \sigma_{lnj}^d[t] \over \sum_{k \in \cN} \hat \sigma_{lnk}^d[t] }.
\end{eqnarray}

\subsubsection{Token Bucket Algorithm}
At each node $n,$ for each previous-hop neighbor $l,$ next-hop neighbor $j$ and each destination $d,$ we maintain a token bucket $r_{lnj}^d.$ At each time slot $t,$ the token bucket is decremented by $\sigma_{lnj}^d[t],$ but cannot go below the lower bound $0:$
\begin{eqnarray*}
r_{lnj}^d[t]=\max\{r_{lnj}^d[t-1]-\sigma_{lnj}^d[t],0\}.
\end{eqnarray*}
When $r_{lnj}^d[t-1]-\sigma_{lnj}^d[t]<0,$ we say $\sigma_{lnj}^d[t]-r_{lnj}^d[t-1]$ tokens (associated with bucket $r_{lnj}^d$) are ``wasted" in slot $t.$ Upon a packet arrival from previous hop $l$ at node $n$ for destination $d$ at slot $t,$ we find the token bucket $r_{lnj*}^d$ which has the smallest number of tokens (the minimization is over next-hop neighbors $j$), breaking ties arbitrarily, add the packet to the corresponding real queue $q_{lnj^*},$ and add one token from the corresponding bucket:
\begin{eqnarray*}
r_{lnj^*}^d[t]=r_{lnj^*}^d[t]+1.
\end{eqnarray*}

\subsection{Extra link Activation}
Like the case without network coding, extra link activation can reduce delays significantly.
As in the case without network coding, we add additional links to the schedule based on the queue lengths at each link. For extra link activation purposes, we only consider point-to-point links and not broadcast. Thus, we schedule additional point-to-point links by giving priority to those links with larger queue backlogs.

\section{Simulations}
\label{sec:simulation}
We consider two types of networks in our simulations: wireline and wireless. Next, we describe the topologies and simulation parameters used in our simulations, and then present our simulation results.

\subsection{Simulation Settings}

\subsubsection{Wireline Setting}

\begin{figure}[h]
    \begin{center}
        \includegraphics[width=70mm]{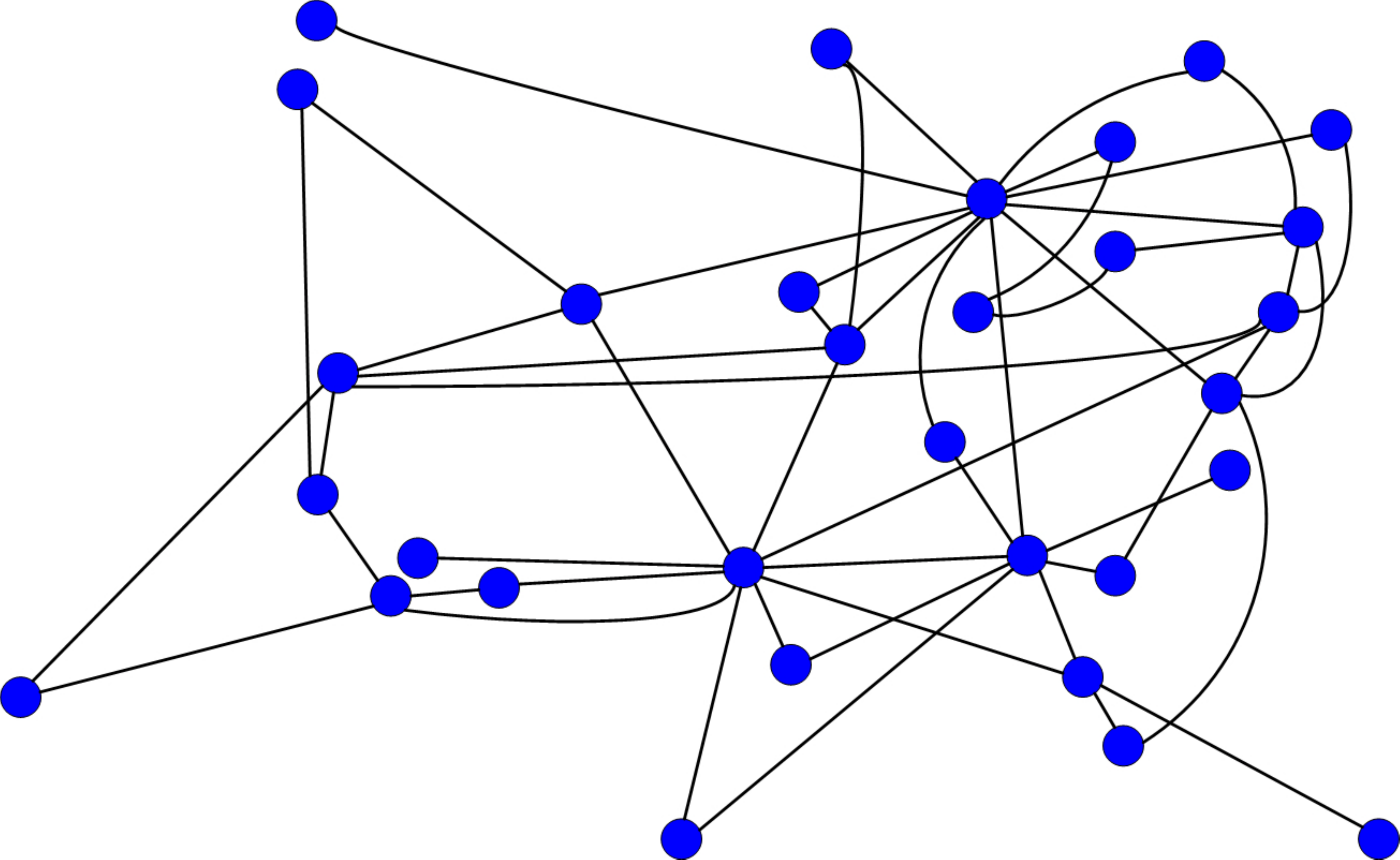}
    \end{center}
        \caption{Sprint GMPLS network topology of North America with $31$ nodes.\cite{sprint}}
        \label{Wireline_network31nodes}
\end{figure}

The network shown in Figure~\ref{Wireline_network31nodes} has $31$ nodes and represents the GMPLS network topology of North America \cite{sprint}. Each link is assume to be able to transmit $1$ packets in each slot. We assume that the arrival process is a Poisson process with parameter $\lambda,$ and we consider the arrivals come within a slot are considered for service at the beginning of the next slot. Once a packet arrives from an external flow at a node $n$, the destination is decided by probability mass function $\hat P_{nd}, d=1,2,...N,$ where $\hat P_{nd}$ is the probability that a packet is received externally at node $n$ destined for $d.$ Obviously, $\sum_{d: d \neq n} \hat P_{nd}=1,$ and $\hat P_{nn}=0.$
The probability $\hat P_{nd}$ is calculated by
\begin{eqnarray*}
\hat P_{nd} = {J_d+J_n \over \displaystyle \sum_{k: k \neq n} (J_k + J_n)},
\end{eqnarray*}
where $J_n$ denotes the number of neighbors of node $n.$ Thus, we use $\hat P_{nd}$ to split the incoming traffic to each destination based on the degrees of the source and the destination.
%
%Because we assume the arrival process is a Poisson process, the arrival process of a flow destined for $d$ at node $n$ is also a Poisson process with parameter $\lambda \bar P_{nd}.$
%

\subsubsection{Wireless Setting}
\begin{figure}[h]
    \begin{center}
        \includegraphics[width=60mm]{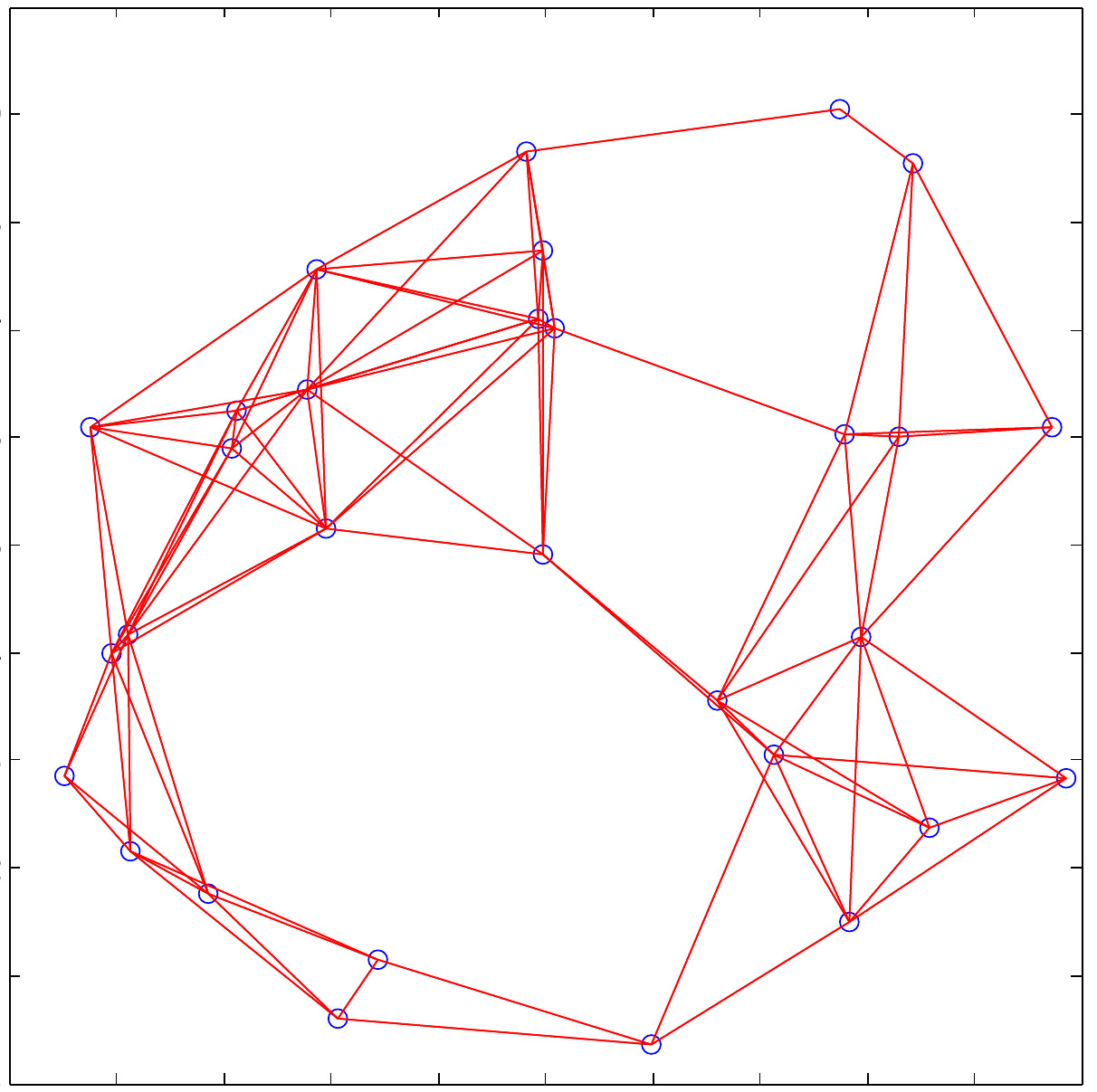}
    \end{center}
        \caption{Wireless network topology with $30$ nodes.}
        \label{Wireless_network30nodes}
\end{figure}

We generated a random network with $30$ nodes which resulted in the topology in Figure~\ref{Wireless_network30nodes}. We used the following procedure to generate the random network: $30$ nodes are placed uniformly at random in a unit square; then starting with a zero transmission range, the transmission range was increased till the network was connected.
We assume that each link can transmit one packet per time slot. We assume a $2$-hop interference model in our simulations. By a $k$-hop interference model, we mean a wireless network where a link activation silences all other links which are $k$ hops from the activated link. The packet arrival processes are generated using the same method as in the wireline case. We simulate two cases given the network topology: the no coding case and the network coding case. In both wireline and wireless simulations, we chose $\beta$ in (\ref{eq: averaging_new}) to be $0.02$.

\subsection{Simulation Results}

\subsubsection{Wireline Networks}

\begin{figure}[h]
    \begin{center}
        \includegraphics[width=80mm]{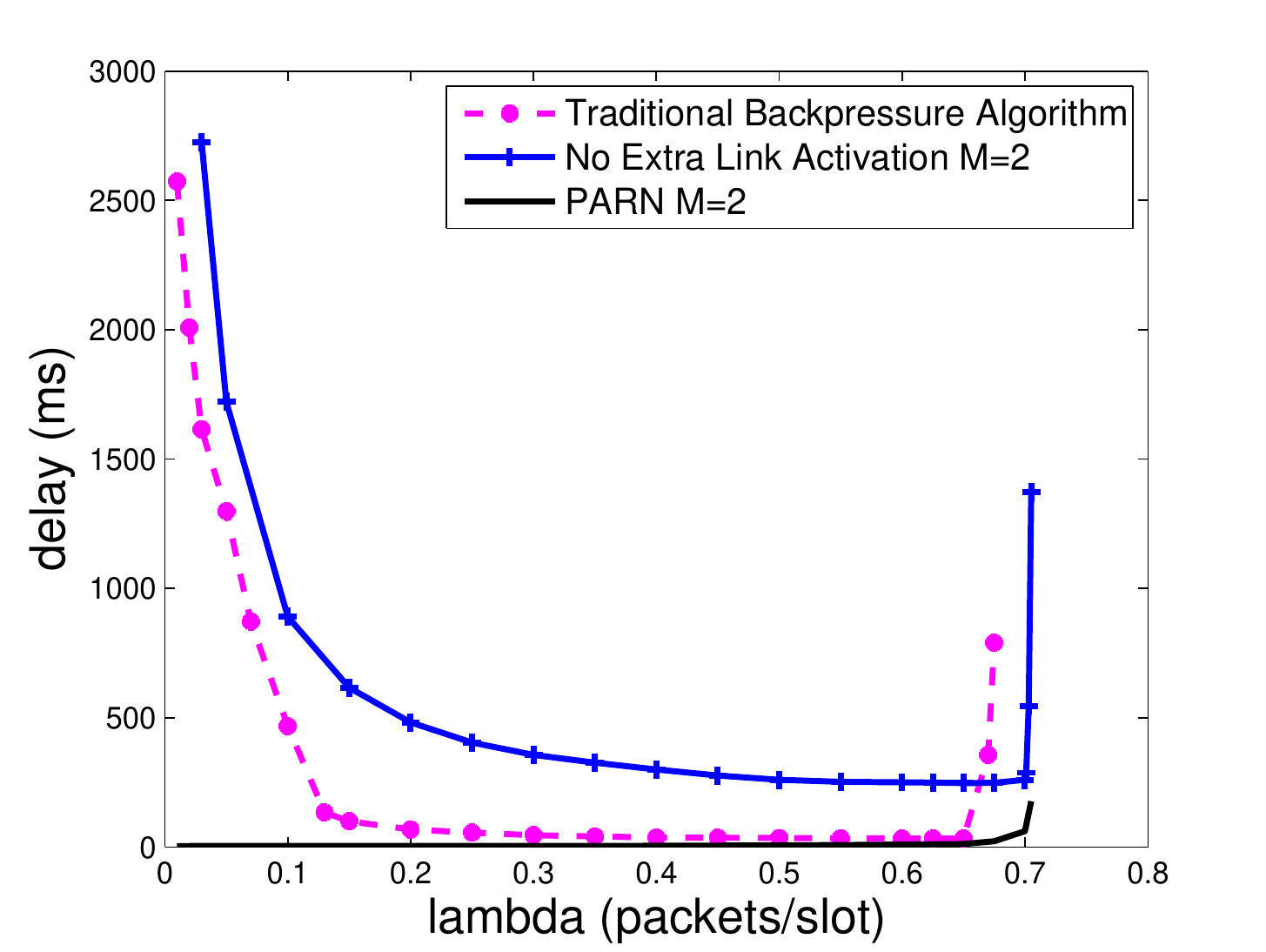}
    \end{center}
        \caption{The impact of the parameter $M$ in Sprint GMPLS network topology.}
        \label{fig:wireline_impact_M}
\end{figure}

First, we compare the performance of three algorithms: the traditional back-pressure algorithm, the basic shadow queue routing/scheduling algorithm without the extra link activation enhancement and PARN. Without extra link activation, to ensure that the real arrival rate at each link is less than the link capacity provided by the shadow algorithm, we choose $\varepsilon=0.02.$ Figure~\ref{fig:wireline_impact_M} shows delay as a function of the arrival rate lambda for the three algorithms. As can be seen from the figure, simply using a value of $M>0$ does not help to reduce delays without extra link activation. The reason is that, while $M>0$ encourages the use of shortest paths, links with back-pressure less than $M$ will not be scheduled and thus can contribute to additional delays.

Next, we study the impact of $M$ on the performance on PARN.

%To emphasize the impact of the parameter $M$ and the extra link activation, we show the delay curves under the backpressure algorithm, the algorithm with parameter $M$ but no extra link activation, and the algorithm with parameter $M$ and  extra link activation in Figure~\ref{fig:wireline_impact_M}. In order to stabilize the wireline network under the algorithm without extra link activation, we choose $\varepsilon$ to be $0.02.$ It shows that the parameter $M$ reduces the delay compared with the back-pressure algorithm. However, without extra link activation, the delay is still high especially in the light traffic region. This is because there is not enough back pressure when the traffic is light. By applying extra link activation, the delay performance outperforms both of the algorithms mentioned previously. It is worth to mention that the throughput region decreases a little bit under the algorithm with parameter $M$ but no extra link activation because of the parameter $\varepsilon.$ Notice that we use probabilistic splitting algorithm in the wireline case because the speed of algorithm is not the bottleneck in the wireline case.

\begin{figure}[h]
    \begin{center}
        \includegraphics[width=80mm]{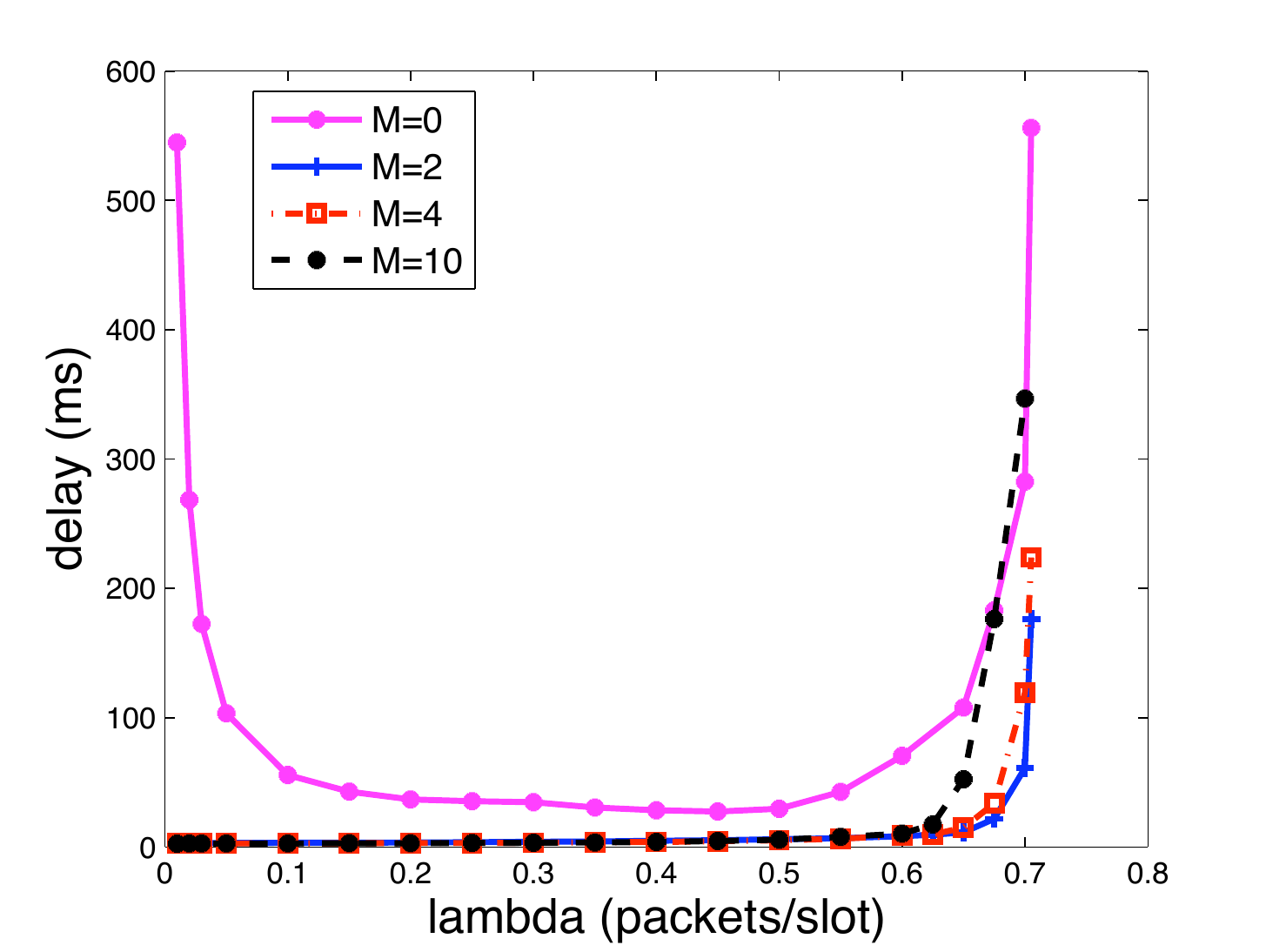}
    \end{center}
        \caption{Packet delay as a function of $\lambda$ under PARN in Sprint GMPLS network topology.}
        \label{fig:wireline_M}
\end{figure}

Figure~\ref{fig:wireline_M} shows the delay performance for various $M$ with extra link activation in the wireline network. The delays for different values of $M$ (except $M=0$) are almost the same in the light traffic region. Once $M$ is sufficiently larger than zero, extra link activation seems to play a bigger role, than the choice of the value of $M,$ in reducing the average delays.

The wireline simulations show the usefulness of the PARN algorithm for adaptive routing. However, a wireline network does not capture the scheduling aspects inherent to wireless networks, which is studied next.

%\subsection{Wireline Simulation Results}

%To check the sensitivity of $M$ in the wireline case, we plot the packet delay as a function of the traffic load for various of $M.$ As shown in Figure~\ref{fig:wireline_M}, the delay performances of different $M$ are almost the same at low traffic. Choosing $M$ to be $4,$ the delay keeps below $8ms$ as long as $\lambda$ is less than $0.6$ packets/slot (equivalently $62.4$ Mbps). Since the average delay of a packet of the shortest path routing is around several milliseconds at low traffic ignoring the queueing delay, this result is really exciting. The delay at low traffic is small because we apply extra link activation.

%We compare PARN with the traditional back-pressure algorithm in Figure~\ref{fig:wireline_compare}. The delay decreases as the traffic load grows at low traffic region and increases suddenly once the traffic load close to the capacity boundary under the traditional back-pressure algorithm. The delay is around hundred or thousand milliseconds depending upon the traffic load of the network under the traditional back-pressure algorithm. Obviously, PARN outperforms the traditional back-pressure algorithm in the whole range of the feasible traffic load. Maintaining a small number of queues at each node is beneficial for the good delay performance. The simulation results also implicitly show that PARN can avoid long path if necessary.

%\subsection{Wireless Simulation Results}

\subsubsection{Wireless Networks}

In the case of wireless networks, even with extra link activation, to ensure stability even when the arrival rates are within the capacity region, we need $\varepsilon > 0.$ We chose $\varepsilon=0.1$ in our simulations due to reasons mentioned in Section~\ref{sec:implement_detail}.

\begin{figure}[h]
    \begin{center}
        \includegraphics[width=80mm]{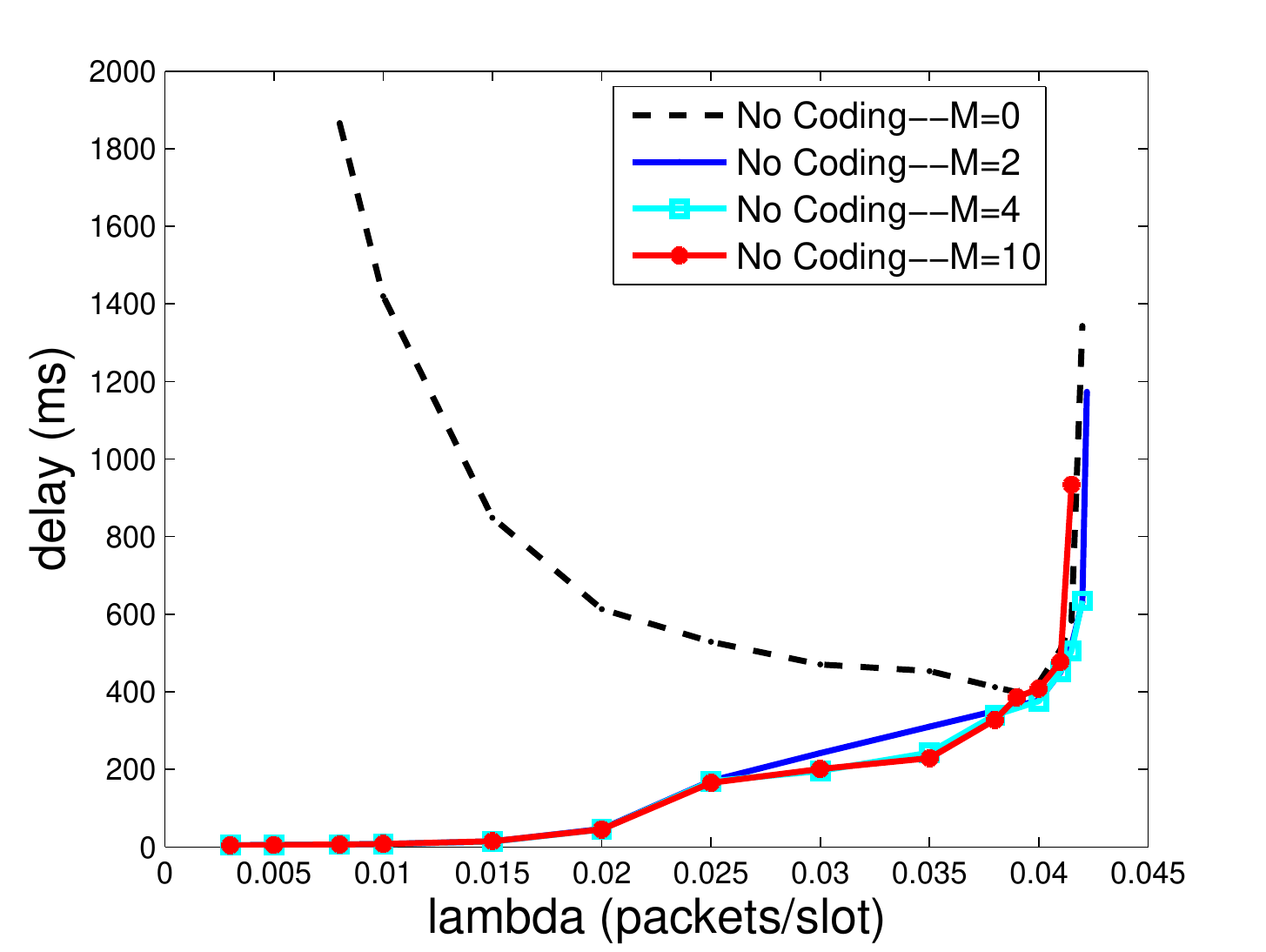}
    \end{center}
        \caption{Packet delay as a function of $\lambda$ under PARN in the wireless network under $2$-hop interference model without network coding.}
        \label{fig:wireless2hop_nocode}
\end{figure}

In Figure~\ref{fig:wireless2hop_nocode}, we study wireless networks without network coding. From the figure, we see that the delay performance is relatively insensitive to the choice of $M$ as long as it is sufficiently greater than zero. The use of $M$ ensures that unnecessary resource wastage does not occur, and thus, extra link activation can be used to decrease delays significantly.

\begin{figure}[h]
    \begin{center}
        \includegraphics[width=80mm]{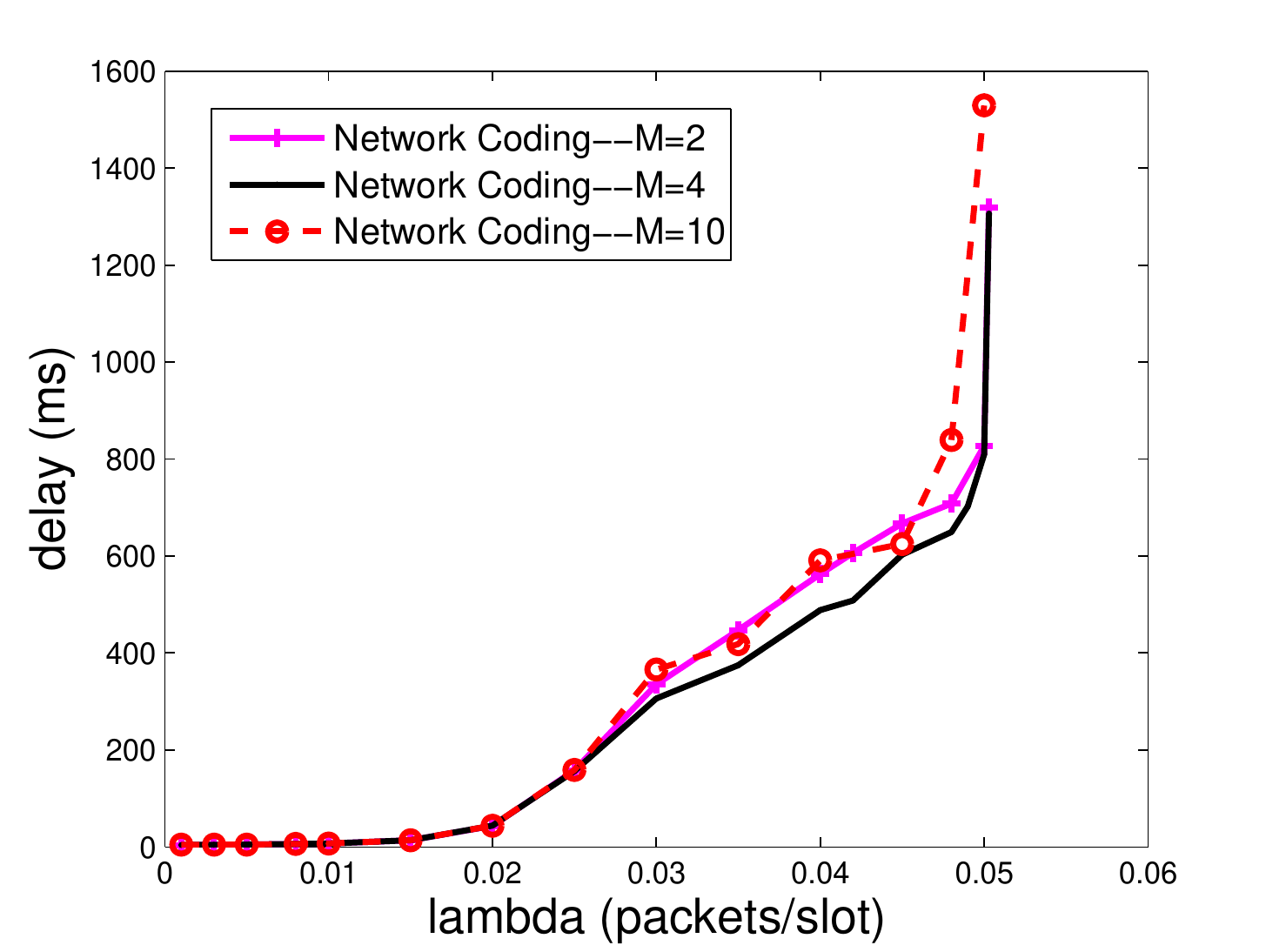}
    \end{center}
        \caption{Packet delay as a function of $\lambda$ under PARN for $M>0$ in the wireless network under $2$-hop interference model with network coding.}
        \label{fig:wireless2hop_netcode}
\end{figure}

\begin{figure}[h]
    \begin{center}
        \includegraphics[width=80mm]{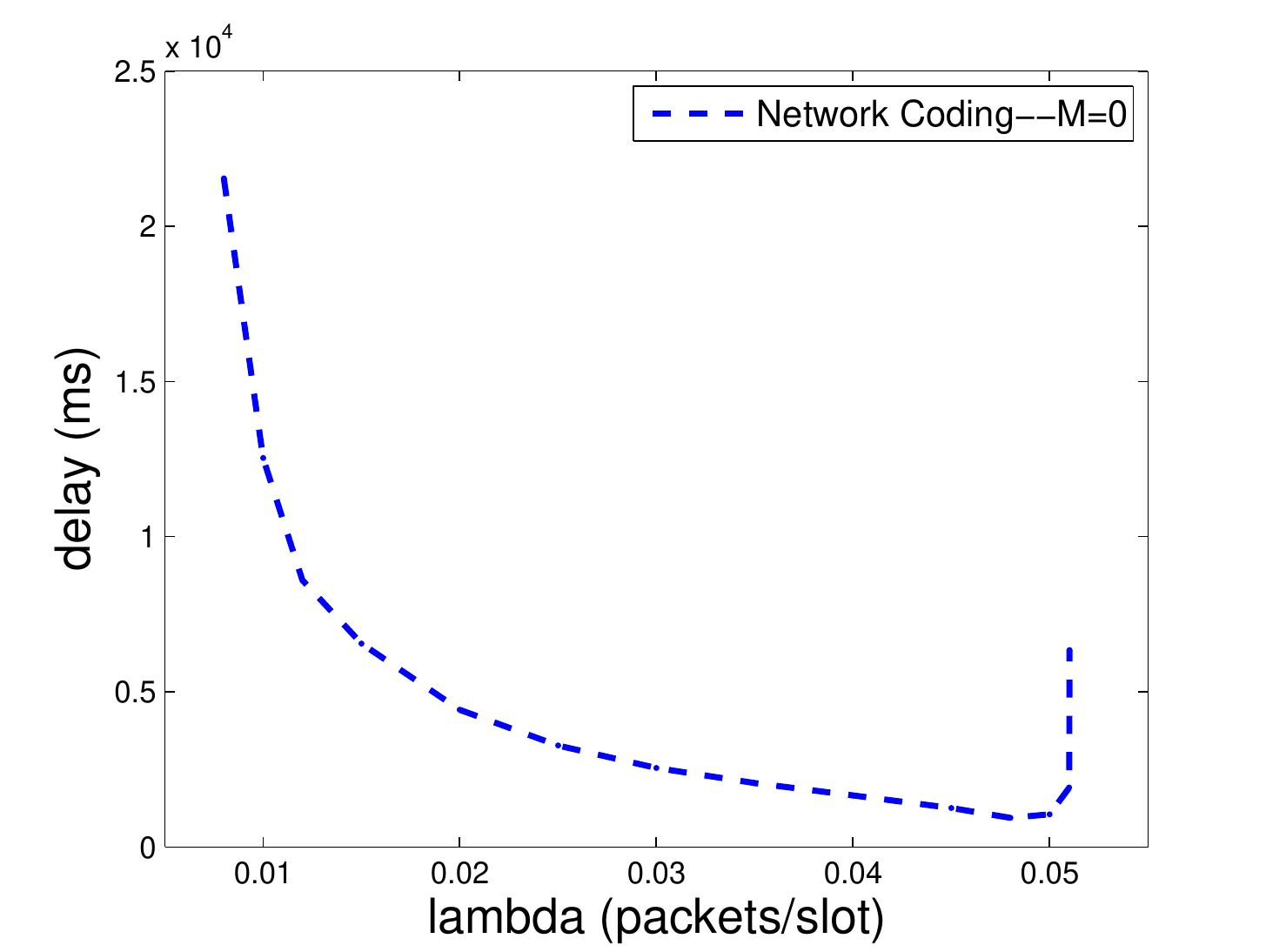}
    \end{center}
        \caption{Packet delay as a function of $\lambda$ under PARN for $M=0$ in the wireless network under $2$-hop interference model with network coding.}
        \label{fig:wireless2hop_netcodeM0}
\end{figure}

In Figures~\ref{fig:wireless2hop_netcode} and~\ref{fig:wireless2hop_netcodeM0}, we show the corresponding results for the case where both adaptive routing and network coding are used. Comparing Figures~\ref{fig:wireless2hop_nocode} and~\ref{fig:wireless2hop_netcode}, we see that, when used in conjunction with adaptive routing, network coding can increase the capacity region. We make the following observation regarding the case $M=0$ in Figure~\ref{fig:wireless2hop_netcodeM0}: in this case, no attempt is made to optimize routing in the network. As a result, the delay performance is very bad compared to the cases with $M > 0$ (Figure~\ref{fig:wireless2hop_netcode}). In other words, network coding alone does not increase capacity sufficiently to overcome the effects of back-pressure routing. On the other hand, PARN with $M>0$ harnesses the power of network coding by selecting routes appropriately.

Next, we make the following observation about network coding. Comparing Figures~\ref{fig:wireless2hop_netcode} and~\ref{fig:wireless2hop_netcodeM0}, we noticed that at moderate to high loads (but when the load is within the capacity region of the no coding case), network coding increases delays slightly. We believe that this is due to fact that packets are stored in multiple queues under network coding at each node: for each next-hop neighbor, a queue for each previous-hop neighbor must be maintained. This seems to result in slower convergence of the routing table.

%We compare the probabilistic splitting algorithm and token bucket algorithm in Figure~\ref{fig:compare_prob_token}.  We choose the wireless network under $1$-hop interference model without network coding for simulations. As shown in Figure~\ref{fig:compare_prob_token}, the token bucket algorithm performs slightly better than the probabilistic splitting algorithm. But the advantage is not remarkable.

\begin{figure}[h]
    \begin{center}
        \includegraphics[width=80mm]{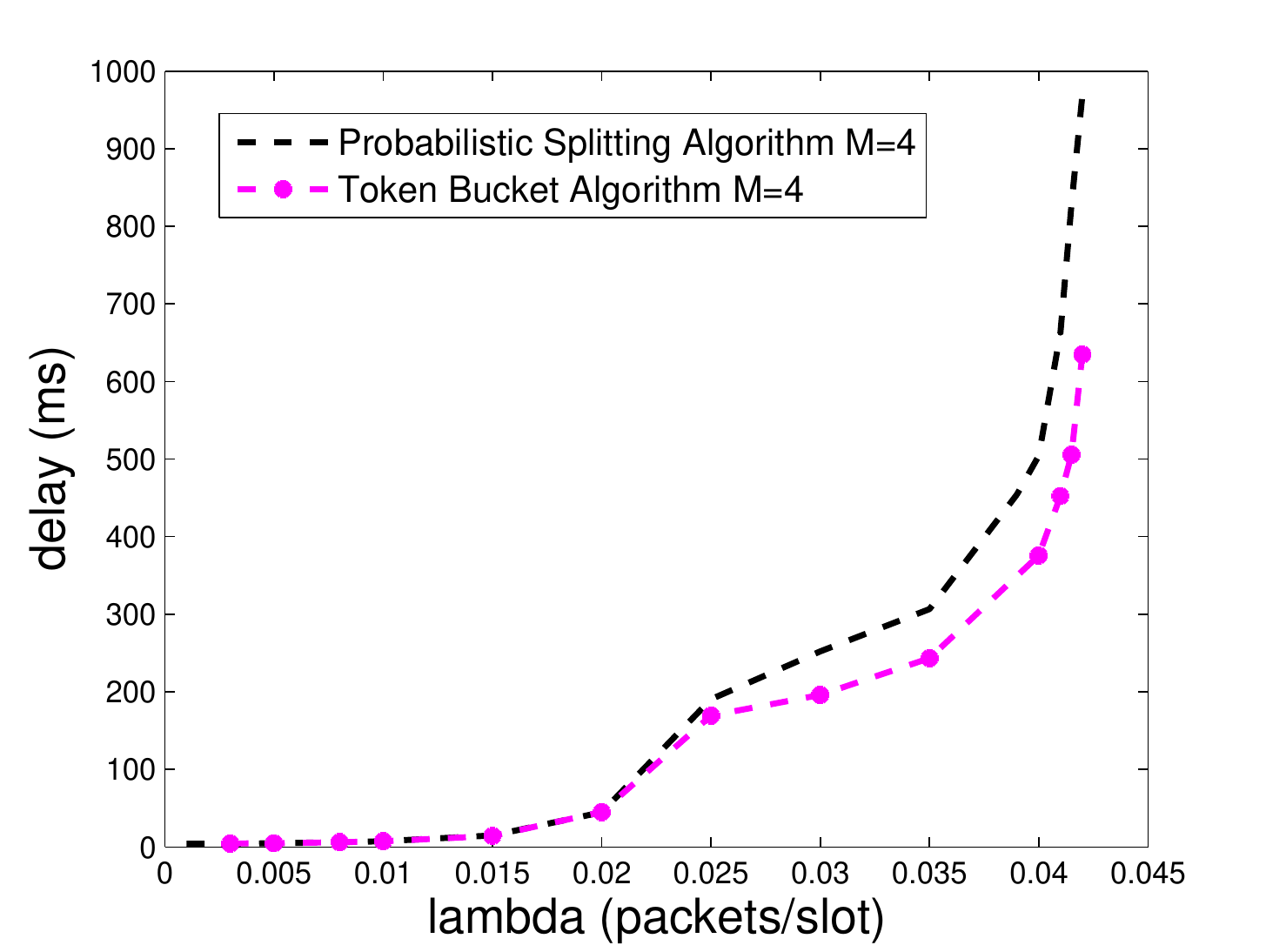}
    \end{center}
        \caption{Comparison of probabilistic splitting and token bucket algorithms under PARN in the wireless network under $2$-hop interference model without network coding.}
        \label{fig:compare_prob_token}
\end{figure}

Finally, we study the performance of the probabilistic splitting algorithm versus the token bucket algorithm. In our simulations, the token bucket algorithm runs significantly faster, by a factor of $2.$ The reason is that many more calculations are needed for the probabilistic splitting algorithm as compared to the token bucket algorithm. This may have some implications for practice. So, in Figure~\ref{fig:compare_prob_token}, we compare the delay performance of the two algorithms. As can be seen from the figure, the token bucket and probabilistic splitting algorithms result in similar performance. Therefore, in practice, the token bucket algorithm may be preferable.

%We also compare the delay performance under different $\varepsilon$'s in Figure~\ref{fig:compare_eps}. The parameter $M$ is chosen to be $2.$ It is shown that by choosing $\varepsilon$ larger, the delay performance will become better. However, larger $\varepsilon$ will shrink the shadow traffic's throughput region more.

%\begin{figure}[h]
%    \begin{center}
%        \includegraphics[width=80mm]{picture/compare_eps.pdf}
%    \end{center}
%        \caption{Compare different $\varepsilon$'s under PARN in the wireless network under $1$-hop interference model without network coding.}
%        \label{fig:compare_eps}
%\end{figure}

\section{Conclusion}
\label{sec:conclusion}

The back-pressure algorithm, while being throughput-optimal, is not useful in practice for adaptive routing since the delay performance can be really bad. In this paper, we have presented an algorithm that routes packets on shortest hops when possible, and decouples routing and scheduling using a probabilistic splitting algorithm built on the concept of shadow queues introduced in \cite{buisristo08, buisristo09}. By maintaining a probabilistic routing table that changes slowly over time, real packets do not have to explore long paths to improve throughput, this functionality is performed by the shadow ``packets." Our algorithm also allows extra link activation to reduce delays. The algorithm has also been shown to reduce the queueing complexity at each node and can be extended to optimally trade off between routing and network coding.

\begin{appendices}

\section{The Stability of the Network Under PARN}
\label{sec:stability-proof}

Our stability result uses the result in \cite{bra96} and relies on the fact that the arrival rate on each link is less than the available capacity of the link.

We will now focus on the case of wireless networks without network coding.

%COM: HERE, ALL WE NEED TO SHOW IS THAT THE ROUTING IS GOVERNED BY IDEAL SPLITTING
%PROBABILITIES, WHICH ARE ALREADY WELL-DEFINED IN THE MAIN TEXT, THEN REAL TRAFFIC LOAD
%OF EACH LINK IS STRICTLY LESS THAN THROUGHPUT PROVEDED BY SHADOW.
%ALL WE NEED TO SAY IS THAT {\boldmath$\lambda = (I - P^T)^{-1} \alpha$}
%HOLDS FOR BOTH SHADOW AND REAL TRAFFIC. THEN, SINCE ``REAL $\alpha$'' IS A SCALED
%DOWN VERSION OF ``SHADOW $\alpha$'', WE ARE DONE. I DO NOT THINK MUCH DETAILS ARE
%NEEDED ...

%\theorem \label{thm:case2} If all shadow queues are stable under the shadow queue algorithm in {\it Case 2}, then all real queues are stable and we call the network is stable in the fluid limit under our algorithm.

%\begin{IEEEproof}
All variables in this appendix are assumed to be average values in the stationary regime of the corresponding variables in the shadow process. Let $\bar \sigma_{nj}^d$ denote the mean shadow traffic rate at link $(nj)$ destined to $d.$ Let $\bar \mu_{nj}$ and $\alpha_n^d(1+\varepsilon)$ denote the mean service rate of link $(nj)$ and the exogenous shadow traffic arrival rate destined to $d$ at node $n.$ Notice that $\varepsilon$ comes from our strategy on shadow traffic. The flow conservation equation is as follows:
\begin{eqnarray}\label{eq:traffic_shadow_eqn}
\alpha_n^d(1+\varepsilon) + \sum_{l: (ln) \in \cL} \bar \sigma_{ln}^d = \sum_{j: (nj) \in \cL} \bar \sigma_{nj}^d, \forall n,d \in \cN.
\end{eqnarray}
The necessary condition on the stability of shadow queues are as follows:
\begin{eqnarray}\label{eq:necessary_fluid_stable}
\sum_{d \in \cN} \bar\sigma_{nj}^d \le \bar\mu_{nj}.
\end{eqnarray}
Since we know that the shadow queues are stable under the shadow queue algorithm, the expression (\ref{eq:necessary_fluid_stable}) should be satisfied.

Now we focus on the real traffic. Suppose the system has an equilibrium distribution and let $\lambda_{nj}^d$ be the mean arrival rate of real traffic at link $(nj)$ destined to $d.$ The splitting probabilities are expressed as follows:
\begin{eqnarray}
P_{nj}^d = {\bar \sigma_{nj}^d \over \sum_{k \in \cN} \bar\sigma_{nk}^d}, \mbox{where}\ \ d \neq n.
\end{eqnarray}
Thus, the mean arrival rates at a link satisfy traffic equation:
\begin{eqnarray}\label{eq:traffic_real_eqn}
\lambda_{nj}^d = \alpha_n^d P_{nj}^d + \sum_{l:(ln) \in \cL} \lambda_{ln}^d P_{nj}^d, \forall (nj) \in \cL, d \in \cN,
\end{eqnarray}
where $d \neq n.$

%Consider link $(nj)$ to be a server with mean service rate $m_j$ determined by the shadow algorithm. Let the traffic destined to $d$ at link $(nj)$ be a type of customer and denote it to be class $(nj,d).$ The term $\alpha_n^d P_{nj}^d$ in (\ref{eq:traffic_real_eqn}) is the exogenous arrival rate of class $(nj, d).$
%$P_{nj}^d$ can be thought of as the transition probability from class $(ln,d)$ to class $(nj,d)$ for any $(ln) \in \cL$ if $l \neq n$ and $d \neq n.$ The vector version of expession (\ref{eq:traffic_real_eqn}) has the form {\boldmath$\lambda = \alpha + P^T \lambda$}. Some rows of matrix {\boldmath$P$} contains all zero entries because we exclude $\lambda_{nj}^n$ from the vector {\boldmath$\lambda$}, and each of the rest of rows contains entries which sum up to $1.$  Hence {\boldmath$P$} is a subprobability matrix and all eigenvalues of it are strictly less than $1.$ Given {\boldmath$\alpha$} and {\boldmath$P$}, we have the unique solution {\boldmath$\lambda = (I - P^T)^{-1} \alpha$}.

%
%The expression (\ref{eq:traffic_real_eqn}) is a special form of the expression (\ref{eq:traffic_eqn}). As long as the traffic intensity at every link $(nj)$ is less than $1,$ the network is stable under our algorithm.
%
The traffic intensity at link $(nj)$ is expressed as:
\begin{eqnarray}\label{eq: traffic_intensity}
\rho_{nj} = {1 \over \bar \mu_{nj}}\sum_{d \in \cN} \lambda_{nj}^d.
\end{eqnarray}
Now we will show $\rho_{nj} <1$ for any link $(nj) \in \cL.$ Let $\lambda_{nj}^d = {\bar \sigma_{nj}^d /(1+\varepsilon)}$ for every $(nj) \in \cL,$ and substitute it into expression (\ref{eq:traffic_real_eqn}). It is easy to check that the candidate solution is valid by using expression (\ref{eq:traffic_shadow_eqn}). From (\ref{eq:necessary_fluid_stable}), the traffic intensity at link $(nj)$ is strictly less than 1 for any link $(nj) \in \cL:$
\begin{eqnarray}
\rho_{nj} ={1 \over \bar \mu_{nj}}\sum_{d \in \cN} \lambda_{nj}^d = {1 \over (1+\varepsilon)\bar \mu_{nj}}\sum_{d \in \cN} \bar\sigma_{nj}^d <1.
\end{eqnarray}
Thus we have shown that the traffic intensity at each link is strictly less than $1.$
%\end{IEEEproof}

The wireline network is a special case of a wireless network. Substitute the link capacity $c_{nj}$ for $\bar\mu_{nj}$ and set $\varepsilon$ to be zero, and stability follows directly.

%\proposition \label{prop:case1} If all shadow queues are stable under the shadow queue algorithm in {\it Case 1}, then all real queues are stable and we call the network is stable in the fluid limit under our algorithm.
%\begin{IEEEproof}
%In {\it Case 1}, the shadow queues and real queues are exactly the same as in {\it Case 2}. In the wireline network, all links can be activated simultaneously. Substitute the link capacity $c_{nj}$ for $\bar\mu_{nj}$ and set $\varepsilon$ to be zero in the proof of Theorem \ref{thm:case2}, and the Proposition \ref{prop:case1} can be proved directly.
%\end{IEEEproof}

%\proposition \label{prop:case3} If all shadow queues are stable under the shadow queue algorithm in {\it Case 3}, then all real queues are stable and we call the network is stable in the fluid limit under our algorithm.
%\begin{IEEEproof}

%\subsection{Wireless Networks with Network Coding}

%This case is analogous to the previous case without network coding. So we omit the proof.

The stability of wireless networks with network coding is similar to the case of wireless network with no coding.

\section{The Back-pressure Algorithm in the Network Coding Case}
\label{sec:find_backpressure_networkcoding}

Given a set of packet arrival rates that lie in the capacity region, our goal is to find routes for flows that use as few resources as possible. Thus, we formulate the following optimization problem for the network coding case.
\begin{eqnarray}
\label{eqn:minhop_netcode}
&\min& \sum_{(nj)\in \bar\cL} \mu_{nj,pp} + \sum_{(n|jl)\in \bar\cL} \mu_{(n|jl)} \\
&s.t.& \mu_{nj,pp}^d + \mu_{nj,broad}^{d} \le \sum_{k} \mu_{njk}^d + \sum_{d'} \sum_{k: k \neq n}\mu_{j|kn}^{d,d'} \nonumber \\
&& \sum_{f \in \cF}x_f I_{\{b(f)=n, e(f)=d\}} \le \sum_{f \in \cF} \sum_{j \in \cN} \mu_{fnj}^d \nonumber
\end{eqnarray}

Let $\{q_{njd}\}$ and $\{q_{0nd}\}$ be the Lagrange multipliers corresponding to the flow conservation constraints in problem (\ref{eqn:minhop_netcode}). Appending the constraints to the objective, we get
\begin{eqnarray} \label{eqn:optimize_netcode}
& &\min_{\mbox{{\boldmath$\mu$}} \in \Lambda'} \sum_{(nj) \in \bar\cL} \mu_{nj,pp} + \sum_{(n|jl)\in \bar\cL} \mu_{n|jl} + \sum_{d}\sum_{(nj)\in \bar\cL} q_{njd}\Bigl[ \nonumber\\
& & \hspace{2mm}  \mu_{nj,pp}^d+ \mu_{nj,broad}^{d}- \sum_{k} \mu_{njk}^d - \sum_{d'} \sum_{k: k \neq n}\mu_{j|kn}^{d,d'} \Bigr]   \\
& & \hspace{2mm} + \sum_{n,d} q_{0nd}\Bigl[ \sum_{f \in \cF}x_f I_{\{b(f)=n, e(f)=d\}} - \sum_{f \in \cF} \sum_{j \in \cN} \mu_{fnj}^d \Bigr] \nonumber \\
& &= \min_{\mbox{{\boldmath$\mu$}} \in \Lambda'} \Bigl( -\sum_{(nj)\in \bar\cL} \sum_{l:(ln)\in \bar\cL}\sum_d \mu_{lnj}^d \bigl(q_{lnd} - q_{njd} -1 \bigr) \nonumber \\
& & \hspace{2mm} -\sum_{(n|jl)\in \bar\cL , j<l} \sum_{d,d'} \mu_{n|jl}^{d,d'}\bigl( q_{lnd} -q_{njd}+q_{jnd'}-q_{nld'}-2 \bigr) \nonumber \\
& & \hspace{2mm} - \sum_{(nj) \in \bar\cL} \sum_d \sum_{f \in \cF} \mu_{fnj}^d \bigl( q_{0nd} -q_{njd}-1 \bigr) \nonumber \\
& & \hspace{2mm} + \sum_{n,d} q_{0nd} \sum_{f \in \cF} x_f  I_{\{b(f)=n, e(f)=d\}} \Bigr). \nonumber
\end{eqnarray}
If the Lagrange multipliers are known, then the optimal {\boldmath$\mu$} can be found by solving
\begin{eqnarray}
\max_{\mbox{{\boldmath$\mu$}}\in \Lambda'} \sum_{(nj)\in \bar\cL} \mu_{nj,pp} w_{nj} + \sum_{(n|jl)\in \bar\cL,j<l} \mu_{n|jl} w_{n|jl}
\end{eqnarray}
where
\begin{eqnarray*}
& & w_{nj} = \displaystyle  \max_d \{ w_{nj}^d \},\\
& & w_{n \mid jl} = \displaystyle \max_{d,d'} \{ w_{n \mid jl}^{d,d'} \},\\
& & w_{n|jl}^{d,d'} = w_{lnj}^d + w_{jnl}^{d'} \\
& & w_{nj}^d = \displaystyle \max_{l: (ln) \in \cL \mbox{\footnotesize{ or }} l=0} w_{lnj}^d  \\
& & w_{lnj}^d  =  q_{lnd} -q_{njd}-1.
\end{eqnarray*}
Similar to the update algorithm of $q_{nd}$ in (\ref{eqn:lag_update}), we can derive the update algorithm to compute $q_{njd}:$
\begin{eqnarray}\label{eqn:lag_update_netcode}
& & q_{njd}[t+1] = \Bigl[ q_{njd}[t] + \frac{1}{M}\bigl(\mu_{nj,pp}^d + \mu_{nj,broad}^{d}  \nonumber \\
& & \hspace{5mm} -\sum_{k} \mu_{njk}^d - \sum_{d'} \sum_{k: k \neq n}\mu_{j|kn}^{d,d'} \bigr) \\
& & \hspace{5mm} + \frac{1}{M}\bigl( \sum_{f \in \cF} x_f  I_{\{b(f)=n, e(f)=d\}} - \sum_{f \in \cF} \sum_{j \in \cN} \mu_{fnj}^d \bigr) \Bigr]^+ \nonumber
\end{eqnarray}

By choosing $\frac{1}{M}$ to be the step-size parameter, $Mq_{njd}$ looks very much like a queue update equation. Replacing $M q_{njd}$ by $p_{njd},$ we get (\ref{eq: Wireless point-to-point weight})-(\ref{eq: shadow queue dynamics2}). It can be shown using the results in \cite{neemodli05, sto05} that the stochastic version of the above equations are stable and that the average rates can approximate the solution to (\ref{eqn:minhop_netcode}) arbitrarily closely.

\end{appendices}

\bibliography{refs}

\begin{thebibliography}{10}

\bibitem{sprint}
Sprint {IP} network performance.
\newblock Available at https://www.sprint.net/performance/.

\bibitem{awelei93}
B.~Awerbuch and T.~Leighton.
\newblock A simple local-control approximation algorithm for multicommodity
  flow.
\newblock In {\em Proc. 34th Annual Symposium on the Foundations of Computer
  Science}, 1993.

\bibitem{bra96}
M.~Bramson.
\newblock Convergence to equilbria for fluid models of {FIFO} queueing
  networks.
\newblock {\em Queueing Systems: Theory and Applications}, 22:5--45, 1996.

\bibitem{BZM06}
A.~Brzezinski, G.~Zussman, and E.~Modiano.
\newblock Enabling distributed throughput maximization in wireless mesh
  networks - a partitioning approach.
\newblock In {\em Proc.~ACM Mobicom}, Sep. 2006.

\bibitem{buisristo09a}
L.~Bui, R.~Srikant, and A.~L. Stolyar.
\newblock A novel architecture for delay reduction in the back-pressure
  scheduling algorithm.
\newblock {\em IEEE/ACM Trans. Networking}.
\newblock Submitted, 2009.

\bibitem{buisristo08}
L.~Bui, R.~Srikant, and A.~L. Stolyar.
\newblock Optimal resource allocation for multicast flows in multihop wireless
  networks.
\newblock {\em Philosophical Transactions of the Royal Society, Ser. A}, 2008.
\newblock To appear.

\bibitem{buisristo09}
L.~Bui, R.~Srikant, and A.~L. Stolyar.
\newblock Novel architectures and algorithms for delay reduction in
  back-pressure scheduling and routing.
\newblock In {\em Proceedings of IEEE INFOCOM Mini-Conference}, April 2009.

\bibitem{cheholowchidoy07}
L.~Chen, T.~Ho, S.~H. Low, M.~Chiang, and J.~C. Doyle.
\newblock Optimization based rate control for multicast with network coding.
\newblock In {\em Proc. IEEE INFOCOM}, Anchorage, Alaska, May 2007.

\bibitem{dimwal06}
A.~Dimakis and J.~Walrand.
\newblock Sufficient conditions for stability of longest-queue-first
  scheduling: Second-order properties using fluid limits.
\newblock {\em Advances in Applied Probability}, June 2006.

\bibitem{effhokim06}
M.~Effros, T.~Ho, and S.~Kim.
\newblock A tiling approach to network code design for wireless networks.
\newblock In {\em Information Theory Workshop}, 2006.

\bibitem{erylun07}
A.~Eryilmaz and D.~S. Lun.
\newblock Control for inter-session network coding.
\newblock In {\em Proceedings of the Workshop on Network Coding, Theory and
  Applications (NetCod)}, January 2007.

\bibitem{erysri05}
A.~Eryilmaz and R.~Srikant.
\newblock Fair resource allocation in wireless networks using
  queue-length-based scheduling and congestion control.
\newblock In {\em Proceedings of IEEE INFOCOM}, 2005.
\newblock Revised version to appear in \emph{IEEE/ACM Transactions on
  Networking}.

\bibitem{erysri06}
A.~Eryilmaz and R.~Srikant.
\newblock Joint congestion control, routing and mac for stability and fairness
  in wireless networks.
\newblock In {\em Proc. International Zurich Seminar on Communications}, 2006.

\bibitem{hovis09}
T.~Ho and H.~Viswanathan.
\newblock Dynamic algorithms for multicast with intra-session network coding.
\newblock {\em IEEE Transactions on Information Theory}, February 2009.

\bibitem{sefmarkoz09}
H.Seferoglu, A.Markopoulou, and U.Kozat.
\newblock Network coding-aware rate control and scheduling in wireless
  networks.
\newblock In {\em Special Session on "Network Coding for Multimedia Streaming",
  ICME}, Cancun, Mexico, June 2009.

\bibitem{joolinshr08}
C.~Joo, X.~Lin, and N.~B. Shroff.
\newblock Understanding the capacity region of the greedy maximal scheduling
  algorithm in multi-hop wireless networks.
\newblock In {\em Proc. IEEE INFOCOM}, 2008.

\bibitem{katrahhukatmedcro06}
S.~Katti, H.~Rahul, W.~Hu, D.~Katabi, M.~Medard, and J.~Crowcroft.
\newblock {XOR}s in the air: Practical wireless network coding.
\newblock In {\em ACM SIGCOMM Computer Communication Review}, volume~36, pages
  243--254, 2006.

\bibitem{lecnisri09}
M.~Leconte, J.~Ni, and R.~Srikant.
\newblock Improved bounds on the throughput efficient of greedy maximal
  scheduling in wireless networks.
\newblock In {\em Proc.~ACM MobiHoc}, 2009.

\bibitem{liboyxia09}
B.~Li, C.~Boyaci, and Y.~Xia.
\newblock A refined performance characterization of longest-queue-first policy
  in wireless networks.
\newblock In {\em Proc.~ACM MobiHoc}, 2009.

\bibitem{linshr04}
X.~Lin and N.~Shroff.
\newblock On the stability region of congestion control.
\newblock In {\em Proceedings of the Allerton Conference on Communications,
  Control and Computing}, 2004.

\bibitem{neemodli05}
M.~J. Neely, E.~Modiano, and C.~Li.
\newblock Fairness and optimal stochastic control for heterogeneous networks.
\newblock In {\em Proceedings of IEEE INFOCOM}, 2005.

\bibitem{neemodroh05}
M.~J. Neely, E.~Modiano, and C.~E. Rohrs.
\newblock Dynamic power allocation and routing for time varying wireless
  networks.
\newblock {\em IEEE Journal on Selected Areas in Communications},
  23(1):89--103, January 2005.

\bibitem{senrayban07}
S.~Banerjee S.~Sengupta, S.~Rayanchu.
\newblock An analysis of wireless network coding for unicast sessions: The case
  for coding-aware routing.
\newblock In {\em Proc. IEEE INFOCOM}, Anchorage, Alaska, May 2007.

\bibitem{sto05}
A.~L. Stolyar.
\newblock Maximizing queueing network utility subject to stability: Greedy
  primal-dual algorithm.
\newblock {\em Queueing Systems}, 2005.

\bibitem{taseph92}
L.~Tassiulas and A.~Ephremides.
\newblock Stability properties of constrained queueing systems and scheduling
  policies for maximum throughput in multihop radio networks.
\newblock {\em IEEE Transactions on Automatic Control}, pages 1936--1948,
  December 1992.

\bibitem{yinshared09}
L.~Ying, S.~Shakkottai, and A.~Reddy.
\newblock On combining shortest-path and back-pressure routing over multihop
  wireless networks.
\newblock In {\em Proceedings of IEEE INFOCOM 2009}, April 2009.

\end{thebibliography}
\bibliographystyle{plain}

\end{document}